\documentclass[ 10pt,twocolumn]{IEEEtran}
\usepackage{float}
\usepackage{multicol}

\usepackage{cite}
\usepackage[cmex10]{amsmath}
\usepackage{amssymb}
\usepackage{amsthm}
\usepackage{mathrsfs}
\usepackage{bm}
\usepackage[mathscr]{eucal}
\usepackage{amssymb,amsmath,amsthm,amsfonts,latexsym}
\usepackage{amsmath,graphicx,bm,xcolor,url}
\usepackage{graphicx}
\usepackage{latexsym}
\usepackage{CJK}
\usepackage{indentfirst}
\usepackage{geometry}
\usepackage{psfrag}
\usepackage{setspace}
\usepackage{algorithmic}
\usepackage{algorithmic, cite}
\usepackage{algorithm}
\usepackage{array}
\usepackage{mdwmath}
\usepackage{mdwtab}
\usepackage{eqparbox}
\usepackage{url}
\usepackage{epstopdf}
\usepackage{epsfig,epsf,psfrag}
\usepackage{fixltx2e}
\usepackage{verbatim}
\usepackage{textcomp}
\usepackage{psfrag} 

\usepackage{caption}
\usepackage{subcaption}

\allowdisplaybreaks[1]

\usepackage{fancyhdr}
\theoremstyle{plain}
\newtheorem{mythe}{Theorem}
\theoremstyle{remark}
\newtheorem{mylem}{Lemma}
\theoremstyle{plain}

\theoremstyle{remark}

\theoremstyle{plain}

\theoremstyle{remark}
\newtheorem{myrem}{Remark}
\theoremstyle{remark}

\theoremstyle{remark}

\theoremstyle{remark}

\theoremstyle{remark}

\theoremstyle{remark}

\catcode`~=11 \def\UrlSpecials{\do\~{\kern -.15em\lower .7ex\hbox{~}\kern .04em}} \catcode`~=13

\allowdisplaybreaks[4]

\newcommand{\calC}{\mathcal{C}}

\newcommand{\calN}{\mathcal{N}}

\newcommand{\ba}{\mathbf{a}}
\newcommand{\bA}{\mathbf{A}}

\newcommand{\bB}{\mathbf{B}}

\newcommand{\bD}{\mathbf{D}}
\newcommand{\be}{\mathbf{e}}
\newcommand{\bE}{\mathbf{E}}

\newcommand{\bg}{\mathbf{g}}
\newcommand{\bG}{\mathbf{G}}

\newcommand{\bH}{\mathbf{H}}

\newcommand{\bI}{\mathbf{I}}

\newcommand{\bn}{\mathbf{n}}
\newcommand{\bN}{\mathbf{N}}

\newcommand{\br}{\mathbf{r}}

\newcommand{\bu}{\mathbf{u}}

\newcommand{\bw}{\mathbf{w}}

\newcommand{\bx}{\mathbf{x}}

\newcommand{\by}{\mathbf{y}}
\newcommand{\bY}{\mathbf{Y}}

\newcommand{\bbE}{\mathbb{E}}

\DeclareMathAlphabet{\mathbsf}{OT1}{cmss}{bx}{n}
\DeclareMathAlphabet{\mathssf}{OT1}{cmss}{m}{sl}

\DeclareSymbolFont{bsfletters}{OT1}{cmss}{bx}{n}
\DeclareSymbolFont{ssfletters}{OT1}{cmss}{m}{n}
\DeclareMathSymbol{\bsfGamma}{0}{bsfletters}{'000}
\DeclareMathSymbol{\ssfGamma}{0}{ssfletters}{'000}
\DeclareMathSymbol{\bsfDelta}{0}{bsfletters}{'001}
\DeclareMathSymbol{\ssfDelta}{0}{ssfletters}{'001}
\DeclareMathSymbol{\bsfTheta}{0}{bsfletters}{'002}
\DeclareMathSymbol{\ssfTheta}{0}{ssfletters}{'002}
\DeclareMathSymbol{\bsfLambda}{0}{bsfletters}{'003}
\DeclareMathSymbol{\ssfLambda}{0}{ssfletters}{'003}
\DeclareMathSymbol{\bsfXi}{0}{bsfletters}{'004}
\DeclareMathSymbol{\ssfXi}{0}{ssfletters}{'004}
\DeclareMathSymbol{\bsfPi}{0}{bsfletters}{'005}
\DeclareMathSymbol{\ssfPi}{0}{ssfletters}{'005}
\DeclareMathSymbol{\bsfSigma}{0}{bsfletters}{'006}
\DeclareMathSymbol{\ssfSigma}{0}{ssfletters}{'006}
\DeclareMathSymbol{\bsfUpsilon}{0}{bsfletters}{'007}
\DeclareMathSymbol{\ssfUpsilon}{0}{ssfletters}{'007}
\DeclareMathSymbol{\bsfPhi}{0}{bsfletters}{'010}
\DeclareMathSymbol{\ssfPhi}{0}{ssfletters}{'010}
\DeclareMathSymbol{\bsfPsi}{0}{bsfletters}{'011}
\DeclareMathSymbol{\ssfPsi}{0}{ssfletters}{'011}
\DeclareMathSymbol{\bsfOmega}{0}{bsfletters}{'012}
\DeclareMathSymbol{\ssfOmega}{0}{ssfletters}{'012}

\newcommand{\tilg}{\widetilde{g}}

\newcommand{\hatbg}{\widehat{\bg}}
\newcommand{\hatbG}{\widehat{\bG}}
\newcommand{\tilbg}{\widetilde{\bg}}

\newcommand{\tilR}{\widetilde{R}}

\newcommand{\bSigma	}{\bm{\Sigma}}

\newcommand{\bPhi}{\bm{\Phi}}

\def\norm#1{\left\| #1 \right\|}
\def\norm2#1{\left\| #1 \right\|_2}
\def\norm22#1{\left\| #1 \right\|_2^2}

\newcommand{\eqa}{\stackrel{(a)}{=}}
\newcommand{\eqb}{\stackrel{(b)}{=}}

\DeclareMathOperator{\diag}{diag}

\newcommand{\qednew}{\nobreak \ifvmode \relax \else
      \ifdim\lastskip<1.5em \hskip-\lastskip
      \hskip1.5em plus0em minus0.5em \fi \nobreak
      \vrule height0.75em width0.5em depth0.25em\fi}

\newcommand{\tilgamma}{\widetilde{\gamma}}

\newcommand{\convergeM}{\stackrel{}{\xrightarrow{\hspace*{0.6cm}}}}

\title{Throughput Optimization for Massive MIMO Systems Powered by Wireless Energy Transfer}

\author{Gang~Yang, Chin~Keong~Ho, Rui~Zhang, and Yong~Liang~Guan  
\thanks{G.~Yang and Y.~L.~Guan are with the School of Electrical and Electronic Engineering, Nanyang Technological University, Singapore (e-mail:\{yang0305, eylguan\}@ntu.edu.sg).} 
\thanks{C. K. Ho is with the Institute for Infocomm Research, A$^\star$STAR, Singapore (e-mail: hock@i2r.a-star.edu.sg). }
\thanks{R.~Zhang is with the Department of Electrical and Computer Engineering, National University of Singapore (e-mail: elezhang@nus.edu.sg). He is also with the Institute for Infocomm Research, A$^\star$STAR, Singapore.}}

\begin{document}
\maketitle 
\vspace{-0.2in}

\begin{abstract}
This paper studies a wireless-energy-transfer (WET) enabled massive multiple-input-multiple-output (MIMO) system (MM) consisting of a hybrid data-and-energy access point (H-AP) and multiple single-antenna users. In the WET-MM system, the H-AP is equipped with a large number $M$ of antennas and functions like a conventional AP in receiving data from users, but additionally supplies wireless power to the users. We consider frame-based transmissions. Each frame is divided into three phases: the uplink channel estimation (CE) phase, the downlink WET phase, as well as the uplink wireless information transmission (WIT) phase. Firstly, users use a fraction of the previously harvested energy to send pilots, while the H-AP estimates the uplink channels and obtains the downlink channels by exploiting channel reciprocity. Next, the H-AP utilizes the channel estimates just obtained to transfer wireless energy to all users in the downlink via energy beamforming. Finally, the users use a portion of the harvested energy to send data to the H-AP simultaneously in the uplink (reserving some harvested energy for sending pilots in the next frame). To optimize the throughput and ensure rate fairness, we consider the problem of maximizing the minimum rate among all users. In the large-$M$ regime, we obtain the asymptotically optimal solutions and some interesting insights for the optimal design of WET-MM system.
\end{abstract}


\section{Introduction}
Recently, far-field wireless energy transfer (WET) has emerged as a promising technology to address energy and lifetime bottlenecks for power-limited devices in wireless networks~\cite{NShinohara11}~\cite{BiZhangMag14}. WET refers to using the radiative electromagnetic (EM) wave emitted from a power transmitter to deliver energy to a power receiver (see~\cite{NShinohara11}~\cite{BiZhangMag14} and references therein). Since EM waves decay quickly over distances, to realize WET in practice, the EM energy needs to be concentrated into a narrow beam to achieve efficient transmission of power, also referred to as {\textit{energy beamforming}}~\cite{MIMOWIPTZhang13}.

Simultaneous wireless information and power transfer (SWIPT) that was proposed in~\cite{MIMOWIPTZhang13}~\cite{Varshney08}, has been extensively studied in literature, since it offers great convenience to mobile users with concurrent data and energy supplies. The authors in~\cite{ZhouZhangHoArchitecture13},~\cite{MIMOWIPTZhang13} studied the performance limits of single-input-single-output (SISO) and  multiple-input-multiple-output (MIMO) SWIPT systems, respectively, and characterized various achievable rate-energy (R-E) trade-offs by practical receiver designs. SWIPT has also been studied in fading channels~\cite{WITOppoEHLiuZhangChua13}, orthogonal frequency division multiplexing (OFDM) systems~\cite{GroverShannonTesla10,KHuangELarsson13,DNgRScholar13,ZhouZhangHo14}, and multiuser channel setups such as broadcast channels~\cite{SWIPTXuZhangTSP14,RandomBFJuZhang13}, relay channels~\cite{SWIPTSimeone12,ANasirAKennedy12}, and interference channels~\cite{SWIPTShenLiChang12,SWIFTInterferClerckx13}. Moreover,~\cite{KHuangVLau13} studied a hybrid network which overlays an uplink cellular network with randomly deployed power beacons that charge users wirelessly, while~\cite{SLeeKHuang13} studied a similar setup in cognitive radio networks with secondary users harvesting wireless power opportunistically from nearby primary users' transmission.

Another emerging trend focuses on the study of using wireless power to support wireless communications, thus forming a wireless powered communication network (WPCN)~\cite{HJuRZhang13}. In a WPCN, an access point (AP) with multiple antennas first transfers energy to multiple single-antenna users via downlink beamforming, then the users use the harvested energy to perform uplink wireless
information transmission (WIT) to the AP. Due to the channel propagation loss, the harvested
energy decays exponentially with respect to the distance between users and the AP. Hence,
the challenge for deploying a WPCN in practice is to ensure that the uplink WIT is feasible when
users are solely powered by downlink WET. Some commercial products on WPCN are currently available~\cite{PowercastURL}. Moreover, several technological advancements provide impetuses to the mass deployment of WPCN in the near future. First, the ongoing deployment of small-cell networks like femtocell~\cite{AndrewsFemtocell13} shortens the energy transfer distance, and thus decreases the channel propagation attenuation significantly. Second, the advancement in low-power
electronics like the ultra-low-power transceiver~\cite{IMECULP2014} is substantially reducing the required uplink transmission power. Third, the use of concurrent downlink WET and uplink WIT was investigated for multiple users in~\cite{KangHoSun14}. Finally, massive antenna arrays~\cite{RusekLarsson13}~\cite{LuLiMMIMO14} with tens to hundreds of elements can be used to form sharp energy beams towards users to achieve a point-to-point WET efficiency of close to one. Massive MIMO has been prototyped and shown to enormously improve the transmission capacity by exploiting its large array gain at the base stations (BSs)~\cite{RusekLarsson13}. However, little is exactly known on its ability of enhancing the efficiency and distance of WET, which is studied in this work.

In a WPCN, the total time duration is divided into the downlink WET and the uplink WIT. Assuming perfect CSI, the single-user scenario was studied in~\cite{XChenCYuen13, ChenWangChen13}. With finite-rate feedback, ~\cite{XChenCYuen13} optimized the time duration for downlink WET to maximize a lower bound on the uplink WIT rate. \cite{ChenWangChen13} maximized the energy efficiency of uplink WIT, by jointly optimizing the time duration and transmit power for downlink WET. Also assuming perfect CSI, the multiuser WPCN was studied in~\cite{HJuRZhang13, LiuZhangChua13}. In particular, ~\cite{HJuRZhang13} considered the scenario in which users use the harvested energy to send independent information to the AP through time division multiple access (TDMA). The sum throughput was maximized subject to user fairness by jointly optimizing the time allocation for downlink WET and uplink WIT. ~\cite{LiuZhangChua13} maximized the minimum throughput between all users and the AP equipped with multiple antennas, by optimizing the downlink energy beamformer, the uplink transmit power, and searching the optimal time allocation for downlink WET.

In practice, perfect CSI at the transmitter is not available due to various factors such as channel estimation error, feedback error, and time-varying channel. The knowledge of accurate CSI is however especially important in a WPCN. More accurate CSI contributes to \emph{both} higher efficiency of energy transfer and higher uplink information rate. Typically, the AP needs to perform channel estimation (CE) first. Although a longer time duration for CE leads to more accurate CSI available at the AP, it reduces the WET and WIT duration, which may lead to less harvested energy and lower throughput. To optimize the throughput, there is thus a design freedom, that is the time spent for CE, WET and WIT.

For a point-to-point WET system, the effect of CE and feedback on energy beamforming was studied in our previous work~\cite{YangHoGuan13} and~\cite{XuZhangTSP14}. In particular, \cite{YangHoGuan13} investigated the dynamic allocation of time resource for CE and energy resource for WET. The optimal preamble length is obtained by solving a dynamic programming problem. The solution is a threshold-type policy that depends only on the channel estimate power, and hence allows a low-complexity WET system to be implemented in practice. \cite{XuZhangTSP14} studied transmit energy beamforming by using one-bit feedback from the energy receiver, to facilitate hardware implementation. Based on the one-bit feedback information, the energy transmitter adjusts transmit beamforming and concurrently obtains improved estimates of the channel to the energy receiver.

In this paper, we consider a WET-enabled massive MIMO system (termed WET-MM), which consists of one hybrid data-and-energy access point (H-AP) with constant power supply and equipped with a large-scale antenna array, and a set of distributed single-antenna users that rely on the wireless power sent from the H-AP for uplink transmission. We assume frame-based transmissions with the time-division-duplexing (TDD) protocol. Each frame is divided into three phases: the uplink CE (i.e., channel estimation) phase, the downlink WET (i.e., wireless energy transfer) phase, as well as the uplink WIT (i.e., wireless information transmission) phase. Firstly, users use a fraction of the harvested energy in the previous frames to send pilots, while the H-AP estimates the uplink channels and obtains the downlink CSI by exploiting channel reciprocity. Secondly, the H-AP transfers wireless energy to all users in the downlink via energy beamforming with appropriately designed weights. Thirdly, the users use the rest of the harvested energy (after reserving energy for the CE in the next frame) to send their independent information to the H-AP simultaneously in the uplink. To optimize the throughput and ensure rate fairness, we consider the problem of maximizing the minimum rate among all users. The design variables are the time allocations for uplink CE and downlink WET (subject to a given total time for CE, WET and WIT of each frame), the energy allocation weights in the downlink WET phase for different users, as well as the fraction of energy used for CE (versus WIT) at each user.

To the best of our knowledge, this paper is the first in the literature to consider the WET-MM system with imperfect CSI. To investigate the optimality of the proposed WET-MM system with imperfect CSI, we compare it to the ideal case with perfect CSI known at the H-AP. Here, we introduce a metric, namely, the massive MIMO degree-of-rate-gain (MM-DoRG), which is defined as the asymptotic scaling order of the uplink rate with respect to $\log M$, i.e.,
\begin{align}
  \kappa \triangleq \lim_{M \rightarrow \infty} \frac{R}{\log{M}}, \label{eq:MMDoRG}
\end{align}
where $M$ is the number of transmit antennas at the H-AP, and $R$ is the data rate that depends on $M$. The definition of MM-DoRG is motivated by two observations. First, the optimal practical system with imperfect CSI achieves a rate that asymptotically scales by a factor of two with respect to $\log M$. Second, the asymptotic rate of a massive MIMO system powered by energy broadcasting without beamforming scales according to $\log M $. These two observations will be shown in Section~\ref{SecAsymAnalysis}. Hence, the MM-DoRG can be interpreted as the proportional gain of the asymptotic rate for a massive MIMO system powered by WET with beamforming, with respect to a massive MIMO system powered by energy broadcasting without beamforming. The terminology of MM-DoRG is similar to the well-known degree-of-freedom (DoF). In this paper, we use a lower bound on the achievable rate for analytical tractability, which is numerically shown to be tight. We focus on the use of zero-forcing (ZF) detection in the uplink WIT phase. We show that the proposed WET-MM system has the following advantages:

\begin{itemize}
  \item In terms of MM-DoRG, the proposed WET-MM system is optimal, as it achieves the same MM-DoRG as the ideal case with perfect CSI, i.e., $\kappa_{\sf{WET-MM}}=\kappa_{\sf{Ideal}}=2$. For a massive MIMO system powered by energy broadcasting without beamforming, the MM-DoRG is one. Energy beamforming is thus shown to be crucial for the proposed WET-MM system to achieve high-efficiency WET and the optimality in terms of MM-DoRG.


  \item The proposed WET-MM system achieves the best possible rate fairness among users, as all users asymptotically achieve a common rate in the large-$M$ regime. This asymptotically solves the problem of rate unfairness due to the ``double near-far'' effect in a WPCN~\cite{HJuRZhang13}.


  \item The proposed WET-MM system is of low complexity and low overhead, requiring only conventional beamforming and detection at the H-AP. Also, there is no overhead (control signaling) required for indicating the WET and WIT operation, since the asymptotically optimal time and energy allocation for all users is derived and then fixed in all frames.
  \item Numerical results corroborate the analysis. It is numerically shown that the asymptotic analysis and the asymptotically optimal solution, although obtained under the large-$M$ assumption, is accurate for $M$ as small as 25. Moreover, to achieve a desired common rate for all users with a given maximal AP-user distance, the proposed WET-MM system is numerically shown to require less antennas at the H-AP, roughly a square root of that required by a massive MIMO system powered by energy broadcasting without beamforming.
\end{itemize}

The rest of this paper is organized as follows: Section~\ref{SystemModel} presents the system model. The problem formulation is then given in Section~\ref{sec:Formulation}. We derive the achievable rate in Section~\ref{Sec:LBRateAnalysis}. The asymptotic analysis is given in Section~\ref{SecAsymAnalysis}, followed by the asymptotically optimal solutions given in Section~\ref{SecAsymOptSolution}. Numerical results are given in Section~\ref{Simulation}. Finally, Section~\ref{Conclusion} concludes the paper.

{\textit{Notation:}} Scalars are denoted by letters (or Greek letters), vectors by boldface lower-case letters, and matrices by boldface upper-case letters. $\bI$ and $\mathbf{0}$ denote an identity matrix and an all-zero vector, respectively, with appropriate dimensions. For a matrix $\bA$ of arbitrary size, $\bA^{\ast}, \bA^{T}, \bA^{H}$ denote the conjugate, the transpose and the conjugate transpose of $\bA$, respectively. For a diagonal matrix $\bD$ of order $K$, $\bD^{\frac{1}{2}}$ denotes the diagonal matrix whose $k$-th diagonal entry is the square root of the $k$-th diagonal entry of $\bD$. $\bbE [\cdot]$ denotes the statistical expectation. The distribution of a circularly symmetric complex Gaussian (CSCG) random vector with mean $\bu$ and covariance matrix $\bSigma$ is denoted by $\calC \calN (\bu,\bSigma)$. ``$\sim$'' stands for ``distributed as". $\| \ba \|_2$ denotes the Euclidean norm of a complex vector $\ba$. $O(\cdot)$ denotes the big-O order. ``$\convergeM$'' denotes the convergence as $M \rightarrow \infty$.

\section{System Model} \label{SystemModel}
We consider a wireless powered communication network (WPCN) consisting of a hybrid data-and-energy access point (H-AP) with $M$ antennas, and $K$ single-antenna users. Each user uses the harvested energy to power its uplink information transmission. We assume that the H-AP and all users are perfectly synchronized and operate with a TDD protocol. We consider frame-based transmissions over flat-fading channels on a single frequency band.

\begin{figure}
\centering
\includegraphics[width=1.03\columnwidth] {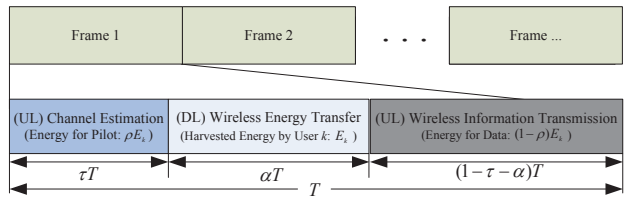}
\caption{Frame Structure}
\label{fig:Fig0}
\end{figure}

As shown in Fig.~\ref{fig:Fig0}, the length of one frame is fixed as $T$ seconds, which is assumed to be less than the coherence interval. Each frame consists of three phases. In the first CE phase of time period $\tau T \ (0 < \tau < 1)$ seconds, the users send orthogonal training pilots, and the H-AP estimates the uplink channels and obtains the downlink CSI by exploiting channel reciprocity. Then in the second WET phase of time period $\alpha T  \ (0 < \alpha < 1)$ seconds, the H-AP delivers energy via beamforming, and the users harvest energy from the received RF signals. In the final WIT phase of the remaining time period $(1-\tau-\alpha) T$ seconds, all users transmit information to the H-AP simultaneously in a space-division-multiplexing-access (SDMA) manner. Clearly, the constraint $0 \leq \tau+\alpha \leq 1$ should be satisfied. For convenience, we normalize
$T = 1$ in the rest of this paper without loss of generality.
Let $\bG = [\bg_1 \ \bg_2 \ \cdots \ \bg_K]$ be the uplink $M \times K$ channel matrix between the H-AP and the $K$ users, i.e., $[\bG]_{mk}= g_{mk} $ is the channel coefficient between the $m$-th antenna of the H-AP and the $k$-th user. We model the matrix $\bG$ as
\begin{align}
  \bG=\bH \bB^{\frac{1}{2}},
\end{align}
where $\bH$ is the $M \times K$ matrix of independent Rayleigh fading coefficients between the H-AP and $K$ users, i.e., $[\bH]_{mk}=h_{mk} \sim \calC \calN (0,1)$, and $\bB$ is a $K \times K$ diagonal matrix, with $[\bB]_{kk} = \beta_k$ denoting the (long-term) path loss of the channel between the H-AP and user $k$ that is assumed to be constant over frames and taken to be known a priori at both the H-AP and user $k$.

\subsection{Uplink Channel Estimation Phase}
In the CE phase, all users simultaneously transmit mutually orthogonal pilot sequences of length $L \ (L\ge K )$ symbols, which allows the H-AP to estimate the channels. In practice, $L$ should be chosen such that the constraint $L T_s \leq \tau$ is satisfied, in which $T_s$ is the sampling period. User $k$ transmits a pilot sequence with power $p_k^{\sf CE}$. Define $\bD=\diag\{L p_{1}^{\sf CE}, \; L p_{2}^{\sf CE},\cdots, L p_{K}^{\sf CE}\}$. The pilot sequences used by $K$ users can be represented by an $L \times K$ matrix $\bPhi \bD^{\frac{1}{2}}$, where $\bPhi$ is of size $L \times K$ and satisfies $\bPhi^H \bPhi=\bI_K$ to preserve orthogonality of the pilots~\cite{NgoLarsson13}. The received signal at the H-AP is thus given by
\begin{align}
  \bY_p = \bG \left(\bPhi \bD^{\frac{1}{2}}\right)^T + \bN,
\end{align}
where $\bN$ is an $M \times L$ matrix with independent and identically distributed (i.i.d.) elements each distributed as $\calC \calN (0,\sigma^2)$. Given $\bY_p$, the minimum mean-square-error (MMSE) estimate of $\bG$, denoted by $\hatbG = [\hatbg_1 \ \hatbg_2 \ \cdots \ \hatbg_K]$, is given by~\cite{SMKayStatisticalSP93}
\begin{align}
  \hatbG &=\bY_p  \bPhi^{\ast} \left( \bB \bD + \sigma^2 \bI_{K} \right)^{-1} \bD^{\frac{1}{2}} \bB. \label{MMSEEstimate2}
\end{align}
Denote the estimation error by $\bE \triangleq \hatbG-\bG$. From the property of MMSE estimation, $\bE$ is independent of $\hatbG$. From~\eqref{MMSEEstimate2}, the elements of the $k$-th column of the matrix $\bE$ are random variables with zero mean and variance
\begin{align}
  \sigma_{e,k}^2 =   \sigma_{e,k}^2 (L, p_k^{\sf CE}) =\frac{\beta_k}{1 + \beta_k L p_{k}^{\sf CE}/\sigma^2}. \label{eq:Errervar1}
\end{align}

\subsection{Downlink Energy Transfer Phase}\label{WPT_Model}
In the downlink WET phase, the $M \times 1$ transmitted signal is given by $\sqrt{p_{\sf dl}} \bw ({\hatbG})$, where $p_{\sf dl}$ is the transmit power for downlink WET, and the beamformer $\bw(\hatbG)$ is to be designed depending on the channel estimate $\hatbG$, subject to $\| \bw ({\hatbG}) \|_2=1$. Let the user noise $n_{0,k} \sim \calC \calN (0, \sigma_0^2)$. Assuming channel reciprocity, the received baseband signal in one symbol period at user $k$ is written as
\begin{align}
  z_k = \sqrt{p_{\sf dl}} \bg_k^H \bw ({\hatbG}) + n_{0,k}.\label{SignalModelMultipleuserWPT}
\end{align}
We assume the energy due to the ambient noise in~\eqref{SignalModelMultipleuserWPT} cannot be harvested. We also assume the energy harvested by user $k$ in one symbol period equals the energy of the equivalent baseband signal in~\eqref{SignalModelMultipleuserWPT} by the law of energy conservation; our results remain valid if a fixed energy loss, or a fixed fraction energy loss, is incurred. Hence, the expected harvested energy by user $k$ is given by
\begin{align}
  Q_k (L, p_k^{\sf CE}, \alpha, \bw({\hatbG})) = \alpha \bbE_{\bG, \hatbG} \left[ p_{\sf dl} \left| \bg_k^H \bw({\hatbG}) \right|^2\right]. \label{EHEnergy0}
\end{align}
We further obtain the asymptotically optimal energy beamformer in the following Lemma~\ref{lemma:AsymOptimalBF}.
\begin{mylem}\label{lemma:AsymOptimalBF}
  As $M \rightarrow \infty$, the asymptotically optimal beamformer that maximizes the harvested energy in~\eqref{EHEnergy0} is a linear combination of the normalized channel estimates between the H-AP and users, i.e.,
\begin{align}
  \bw (\hatbG) = \sum_{k=1}^K \sqrt{\xi_k} \frac{\hatbg_k}{\| \hatbg_k \|_2}, \label{eq:beamformerMU}
\end{align}
where the weights $\xi_k$'s are subject to $\sum_{k=1}^K \xi_k=1$.
\end{mylem}
\begin{IEEEproof}
  See proof in Appendix~\ref{AppendixEnergy_generalBF}.
\end{IEEEproof}
The weights $\xi_k$'s represent the energy allocation for downlink WET among users, which will be optimized later in Section~\ref{SecAsymOptSolution}. In the sequel, we use $Q_k(L, p_k^{\sf CE}, \alpha, \xi_k)$ to denote the harvested energy by user $k$, since the beamformer in~\eqref{eq:beamformerMU} depends on only the weights $\xi_k$'s, given $\hatbG$.

We assume that users have infinite battery storage for storing the harvested energy. Note that the energy used for pilot transmission is drawn from the harvested energy in previous frames. At steady state, we assume a fraction $\rho \ (0 < \rho <1)$ of the harvested energy $Q_k(L, p_k^{\sf CE}, \alpha, \xi_k)$ is used by users to send pilots. The amount of pilot energy affects the accuracy of CSI and thus the efficiency of downlink WET. Under this energy-splitting scheme, the harvested energy for user $k$, denoted as $E_k(\alpha, \rho, \xi_k)$, will be derived in Section~\ref{Sec:LBRateAnalysis}.

\subsection{Uplink Data Transmission Phase}
In the uplink WIT phase, the received baseband signal vector at the H-AP is given by
\begin{align} \label{eq:uplinkModel}
  \by= \bG \bx + \bn,
\end{align}
where $\bx=[x_1 \ x_2 \ \cdots \ x_K]^T$ with $x_k = \sqrt{p_k} s_k$, where $p_k$ is the transmit power of user $k$, and $s_k \sim \calC \calN(0, 1)$ is its information-carrying signal that is independent of signals of other users. Here, the noise vector $\bn \sim \calC \calN(\mathbf{0}_M, \sigma^2 \bI_M)$.

Taking into account of the energy consumption for uplink pilots, the energy left for uplink transmission is $(1-\rho) E_k (\alpha, \rho, \xi_k)$. Thus, the transmit power is given by
\begin{align}
  p_k &= p_k (\tau, \alpha, \rho, \xi_k) = \frac{(1-\rho) E_{k} (\alpha, \rho, \xi_k)}{1-\tau-\alpha}. \label{uplinkTXPower}
\end{align}

In the uplink WIT phase, all users simultaneously transmit to the H-AP. The H-AP adopts a linear detector $\bA=[\ba_1 \; \ba_2\; \cdots \; \ba_K]$ to detect the information for all users. Specifically, a ZF or MRC detector which performs well for large $M$~\cite{RusekLarsson13}, is given by
\begin{align}
  \bA=\left\{ \begin{array}{cl}
  \hatbG \left( \hatbG^H \hatbG \right)^{-1}, &\mbox{for ZF}\\
  \hatbG, \qquad &\mbox{for MRC}. \\
  \end{array}
  \right. \label{ZFMRCdetector}
\end{align}

From~\eqref{eq:uplinkModel}, the signal after using the detector $\bA$ is obtain as $\br=\bA^H \bG \bx + \bA^H \bn$.
In particular, the detected signal associated with user $k$, denoted by $r_k$, is written as
\begin{align}
  r_k &=\sqrt{p_k} \ba_k^H \hatbg_k s_k + \sum \limits_{i=1, i \neq k}^K \sqrt{p_i} \ba_k^H \hatbg_i s_i \nonumber \\
  &\quad \;\; - \sum \limits_{i=1}^K \sqrt{p_i} \ba_k^H \be_i s_i + \ba_k^H \bn,\label{eq:detectedsignal}
\end{align}
where $\ba_i$, $\hatbg_i$, and $\be_i$ are the $i$-th column of $\bA$, $\hatbG$ and $\bE$, respectively. The last three terms in~\eqref{eq:detectedsignal} are treated as interference and noise, and assumed to be Gaussian distributed to maximize entropy. Hence, the achievable rate of the uplink transmission from user $k$ is given by
\begin{align}
  R_k = R_k (\tau, \alpha, \rho, \xi_k) &\triangleq (1 \!-\! \tau \!-\! \alpha) \bbE \big[ \log \left( 1 \!+\! \gamma_k \right) \big],
\end{align}
where the signal-to-interference-plus-noise-ratio (SINR)
\begin{align}
\! \gamma_k = \frac{p_k  |\ba_k^H \hatbg_k |^2}{\sum \limits_{i=1, i \neq k}^K p_i |\ba_k^H \hatbg_i |^2 + |\ba_k^H \ba_k | \sum \limits_{i=1}^K p_i \sigma_{e,i}^2 + |\ba_k^H \ba_k| \sigma^2 }.\nonumber
\end{align}
The exact expression for the rate $R_k$ is not analytically tractable. Instead, we use an analytically tractable lower bound on the rate achieved by using the harvested energy.
By utilizing the convexity of the function $\log \left( 1+ {1}/{x}\right)$ and Jensen's inequality, the rate $R_k$ is lower-bounded (see the same deviation as in Section III-A2 in~\cite{NgoLarsson13}) as
\begin{align}
  R_k \geq \tilR_k 
  &\triangleq (1-\tau-\alpha) \log \left( 1+ \tilgamma_k \right), \label{eq:RateLB}
\end{align}
where the SINR $\tilgamma_k$ is given by the expression
\begin{align}
  &\left(\! \bbE \left[ \!\frac{\sum \limits_{i \neq k} p_i |\ba_k^H \hatbg_i |^2 \!+\! |\ba_k^H \ba_k | \sum \limits_{i=1}^K p_i \sigma_{e,i}^2 \!+\! |\ba_k^H \ba_k| \sigma^2 }{p_k  |\ba_k^H \hatbg_k |^2} \!\right] \!\right)^{-1}. \nonumber
\end{align}
\noindent {We shall observe numerically in Section~\ref{Simulation} that the lower bound in~\eqref{eq:RateLB} is tight for massive MIMO systems. In the sequel, we use $\tilR_k$ as the achievable rate.}

\subsection{Ideal Case and OP-MM System} \label{Sec:Benchmark}
To verify the optimality, we shall compare the proposed WET-MM system with the ideal case, in which the H-AP is assumed to have perfect CSI for downlink energy beamforming and uplink information decoding. The CE phase in Fig.~\ref{fig:Fig0} is thus removed, i.e., $\tau=0, \rho=0, \sigma_{e,k}^2=0, \; \forall k$. Clearly, the WET-MM system with imperfect CSI cannot perform better than the ideal case.

To investigate the performance improvement by introducing energy beamforming in the WET-MM system, we also consider the massive MIMO system powered by energy broadcasting without beamforming, namely the omnidirectional powering massive MIMO (OP-MM) system. The energy broadcasting is being widely used in commercial RF-energy-harvesting products~\cite{PowercastURL,IMECULP2014}. The OP-MM system is of low complexity. The H-AP first broadcasts energy in all directions, i.e., the weight vector $\bw = [1 \; 1 \; \cdots \; 1]^T / \sqrt{M}$. Then it performs uplink CE, followed by uplink WIT. A fraction $\rho$ of the harvested energy is used for CE, with the rest of the energy used for WIT. Hence, channel estimation is employed only for data detection, but not for WET.

\section{Problem Formulation}~\label{sec:Formulation}
Before formulating the problem, we first present the fairness issue in a WPCN. Due to the distance-dependent signal attenuation, the users that are far from the H-AP harvest less energy, and moreover, they have to perform uplink transmission using higher power to overcome the propagation. This is referred as the ``double near-far'' effect~\cite{HJuRZhang13}, which intuitively leads to user unfairness, with significantly lower rates achieved by far users. In our framework, the rates of users can be balanced by adjusting the energy allocation weights $\bm{\xi}=[\xi_1 \ \xi_2 \ \cdots \xi_K]^T$ of the beamformer in~\eqref{eq:beamformerMU}.

To optimize the throughput and achieve user fairness, we maximize the minimum rate for all users, by optimizing over the CE time $\tau$, the WET time $\alpha$, the energy-splitting fraction $\rho$, and the energy allocation vector $\bm{\xi}$. We have the following problem formulation
\begin{subequations}
\label{eq:optimP1}
\begin{align}
   \mathrm{(P_1)}  \quad \max_{\tau, \; \alpha, \; \rho, \; \bm{\xi}} \quad &\min_{1 \leq k \leq K} \tilR_k (\tau, \alpha, \rho, \xi_k) \label{eq:ObjMu} \\
    \quad \text{s. t.} \quad \ \
        & \sum \limits_{k=1}^K \xi_k=1 \label{eq:SumXi0} \\
        & 0 \leq \tau + \alpha \leq 1 \label{eq:constT}\\
    & \tau \geq 0, \ \alpha \geq 0, \ \xi_k \geq  0, \ \forall k \label{eq:Nonnega2} \\
    & 0 \leq \rho \leq 1, \label{eq:RangeRho}
\end{align}
\end{subequations}
where~\eqref{eq:SumXi0} is due to the power constraint at the H-AP for energy beamforming, and~\eqref{eq:constT} is due to the frame duration constraint. The achievable rate $\tilR_k (\tau, \alpha, \rho, \xi_k) $ is given in~\eqref{eq:RateLB}.

For fixed $M$, solving Problem ($\mathrm{P_1}$) is challenging, because of the nonlinear relationship between the information rate and the variables $\tau, \alpha, \rho$ and $\bm{\xi}$. In our system model, the variables are coupled in the uplink CE, downlink WET and uplink WIT phases as follows:

\begin{itemize}
  \item The CE time $\tau$ and the fraction $\rho$ of the harvested energy used for pilots affect the CSI accuracy.

  \item The CSI accuracy affects the amount of harvested energy in the downlink WET phase, the signal power and the interference power for data detection in the uplink WIT phase.
  \item Besides the CSI accuracy, the amount of harvested energy also depends on the energy allocation weights $\bm{\xi}$ and the WET time $\alpha$.
  \item Moreover, the interference from other users depends on the energy harvested by them, the time allocation $\tau, \alpha$, as well as the energy-splitting fraction $\rho$.
\end{itemize}

Nevertheless, it is possible to obtain interesting asymptotical solutions and insights in the large-$M$ regime. In the sequel, we obtain analytical expression for the achievable rate in Section~\ref{Sec:LBRateAnalysis}, followed by the asymptotic analysis in Section~\ref{SecAsymAnalysis}, and then obtain the asymptotically optimal solutions in Section~\ref{SecAsymOptSolution}.

\section{Analysis on Achievable Rate}\label{Sec:LBRateAnalysis}
Before deriving the achievable rate in Section~\ref{subsecLBRate}, we first obtain the harvested energy.
\subsection{Harvested Energy} \label{subsecHarvestedEnergy}
Using the beamformer in~\eqref{eq:beamformerMU}, the harvested energy is derived in the following lemma.
\begin{mylem}\label{lemma:HarvestedEnergy0}
With the beamformer in~\eqref{eq:beamformerMU}, the expected harvested energy by user $k$ is given by
\begin{align}
  \! Q_{k} (L, p_k^{\sf CE}, \alpha, \xi_k) &= \alpha p_{\sf dl} \xi_k \beta_k M \left[ 1 \!-\! \frac{(M \!-\!1)\sigma^2}{M (\beta_k L p_{k}^{\sf CE} +\sigma^2)}\right] \nonumber \\
  &\quad + \alpha p_{\sf dl} \beta_k (1-\xi_k). \label{HarvestedEnergy0}
\end{align}
\end{mylem}
\begin{IEEEproof}
  See Appendix~\ref{AppendixEnergy}.
\end{IEEEproof}
The first term in~\eqref{HarvestedEnergy0} is the harvested energy from the beam directed toward user $k$, while the second term represents the energy harvested from beams directed toward other users but still harvested by user $k$. The expected harvested energy $Q_{k} (L, p_k^{\sf CE}, \alpha, \xi_k)$ increases, as either the (discrete) pilot length $L$ or the pilot power $p_k^{\sf CE}$ increases.


The term $L p_k^{\sf CE}$ in~\eqref{HarvestedEnergy0} represents the energy used by user $k$ for pilot transmission. By assumption, a fraction $\rho$ of the expected harvested energy is used for pilots. With this energy-splitting scheme, we further obtain the harvested energy and the variance of channel estimation error in Lemma~\ref{lemma:HarvestedEnergy}.
\begin{mylem}\label{lemma:HarvestedEnergy}
With the beamformer in~\eqref{eq:beamformerMU} and using the fraction $\rho$ of the expected harvested energy for pilots, the harvested energy by user $k$ is
\begin{align}
  \! \! E_{k} (\alpha, \rho, \xi_k) &\!=\! \frac{g_{k} (\alpha, \rho, \xi_k) \!+\! \sqrt{g_{k}^2 (\alpha, \! \rho, \! \xi_k) \!+\! \frac{4 \alpha p_{\sf dl} \sigma^2}{\rho}}}{2}, \label{HarvestedEnergy1}
\end{align}
where the function
\begin{align}
  g_{k} (\alpha, \rho, \xi_k) = \alpha p_{\sf dl} \beta_k \left( \xi_k (M-1) + 1 \right) -\frac{\sigma^2}{\beta_k \rho}. \nonumber
\end{align}
Moreover, the variance of the channel estimation error of user $k$ is rewritten as
\begin{align}
  \sigma_{e,k}^2 =\sigma_{e,k}^2 (\alpha, \rho, \xi_k) &=\frac{\beta_k \sigma^2}{\beta_k \rho E_{k} (\alpha, \rho,\xi_k) + \sigma^2}. \label{ErrorVariance}
\end{align}
\end{mylem}

\begin{IEEEproof}
Replacing $L  p_{k}^{\sf CE}$ in~\eqref{HarvestedEnergy0} by $\rho Q_k(L, p_k^{\sf CE}, \alpha, \xi_k)$, we obtain the equation with $Q_k (L, p_k^{\sf CE}, \alpha, \xi_k)$ in both sides, which can be written as a quadratic equation in $Q_k(L, p_k^{\sf CE}, \alpha, \xi_k)$. Solving for $Q_k(L, p_k^{\sf CE}, \alpha, \xi_k)$ and discarding the negative solution, we obtain the harvested energy for user $k$ in~\eqref{HarvestedEnergy1}. From~\eqref{eq:Errervar1}, the variance of channel estimation error is obtained as in~\eqref{ErrorVariance}.
\end{IEEEproof}

\subsection{Achievable Rate for WET-MM System with Imperfect CSI} \label{subsecLBRate}
For ZF detection, the detector is $\bA=\hatbG \left( \hatbG^H \hatbG \right)^{-1}$, with $\ba_i^H \hatbg_j$ equals 1 if $i=j$ and equals 0 otherwise. The achievable rate is further obtained in the following lemma.
\begin{mylem}\label{LemmaZF}
  With MMSE channel estimate $\hatbG$, the harvested energy $E_k(\alpha, \rho, \xi_k)$ in~\eqref{HarvestedEnergy1}, and $M \geq K+1$, an achievable uplink rate of user $k$ for ZF detection is given by
  \begin{align}
    \tilR_k^{\sf ZF}
    &=(1-\tau-\alpha) \log \left( 1+ \tilgamma_k \right), \label{eq:RateLBZF}
  \end{align}
    where the SINR
  \[\tilgamma_k \!=\! \frac{(M-K) \beta_k^2 \rho E_k (\alpha, \rho,\xi_k)}{\sigma^2 \big( \!\beta_k \rho \!+\! \frac{\sigma^2}{E_k(\alpha, \rho, \xi_k)} \! \big) \Big(\! \frac{1-\tau-\alpha}{1-\rho} \!+\! \sum \limits_{i=1}^K \frac{\beta_i E_i (\alpha, \rho, \xi_i)}{\beta_i \rho E_i (\alpha, \rho, \xi_k) +\sigma^2} \!\Big)},
  \]
\end{mylem}

\begin{IEEEproof}
Following similar steps as in the proof for Proposition 7 in~\cite{NgoLarsson13}, we have
\begin{align}
    &\tilR_k^{\sf ZF} = (1-\tau-\alpha) \cdot \label{eq:RateLBZF0} \\
    &\; \log \left(\! 1 \!+\! \frac{p_k (\tau, \alpha, \rho, \xi_k) (M \!-\! K) \big(\beta_k \!-\! \sigma_{e,k}^2 (\alpha, \rho, \xi_k) \big)}{\sigma^2 + \sum \limits_{i=1}^K p_i (\tau, \alpha, \rho, \xi_i) \sigma_{e,i}^2 (\alpha, \rho, \xi_i) } \! \right). \nonumber
\end{align}
The result in~\eqref{eq:RateLBZF} is thus obtained, by substituting the error variance in~\eqref{ErrorVariance} into~\eqref{eq:RateLBZF0}.
\end{IEEEproof}

For MRC detection, from~\eqref{ZFMRCdetector}, we have $\bA=\hatbG$. The achievable rate is obtained in Lemma~\ref{LemmaMRC}.
\begin{mylem}\label{LemmaMRC}
With MMSE channel estimate $\hatbG$, MRC detection and $M \geq 2$, an achievable UL rate of user $k$ is given by
  \begin{align}
    \tilR_k^{\text{MRC}}
    &=(1-\tau-\alpha) \log \left( 1+ \tilgamma_k^{\text{MRC}} \right), \label{eq:RateLBMRC}
  \end{align}
  where the SINR
  \begin{align}
  &\tilgamma_k^{\text{MRC}} = \nonumber \\
  &\frac{(M-1) \beta_k^2 \rho E_k (\alpha, \rho, \xi_k)}{\big(\beta_k \rho +\frac{\sigma^2}{E_k(\alpha, \rho, \xi_k)}\big) \Big( \frac{\sigma^2 (1 \!- \! \tau \!-\! \alpha)}{1 -\rho} \!+\! \sum \limits_{i \neq k} \beta_i E_i(\alpha, \rho, \xi_i) \Big) \!+\! \beta_k \sigma^2}, \nonumber
  \end{align}
where the harvested energy $E_k(\alpha, \rho, \xi_k)$ is given by~\eqref{HarvestedEnergy1}.
\end{mylem}

\begin{IEEEproof}
  See Appendix~\ref{AppendixMRC}.
\end{IEEEproof}

For ZF detection, the interference in~\eqref{eq:RateLBZF} is due to only the channel estimation error, while for MRC detection, the interference component in~\eqref{eq:RateLBMRC} comes from both the channel estimation error and the multi-user interference (MUI). Comparatively, MRC detection has lower complexity of computation, since all antennas independently apply a matched filter $\hatbg_k$ to maximize the signal power for user $k$. 



\subsection{Achievable Rate for Ideal Case and OP-MM System} \label{SecLBBenchmarks}
\subsubsection{Ideal Case}
With perfect CSI, i.e., $\sigma_{e,k}^2=0$, the ideal case is the special case for the analysis in Sections~\ref{subsecHarvestedEnergy} and~\ref{subsecLBRate}. With the beamformer in~\eqref{eq:beamformerMU}, the harvested energy by user $k$ is given by
\begin{align}
E_{k} (\alpha, \xi_k) = \alpha p_{dl} \xi_k \beta_k M + \alpha p_{dl} \beta_k (1-\xi_k). \label{eq:EnergyIdealCase}
\end{align} 
Moreover, the achievable rate is given by 
\begin{align}
  &\tilR_{k,\text{Ideal}} =\tilR_{k,\text{Ideal}} (\alpha, \xi_k) \label{eq:RateLBMRCZFPerfectCSI0} \\
  &= \left\{\! \! \!
  \begin{array}{cl}
    (1 \!- \! \alpha) \log \left( 1 \!+\! \frac{E_k (\alpha, \xi_k) (M-K) \beta_k}{(1-\alpha) \sigma^2} \right), \! \!&  \text{for ZF} \\
    (1 \!-\! \alpha) \log \left( 1 \!+\! \frac{E_k (\alpha, \xi_k) (M-1) \beta_k}{\sum \limits_{i \neq k} E_i (\alpha, \xi_i) \beta_i + (1-\alpha) \sigma^2} \right), \! \!&  \text{for MRC.}
  \end{array}
  \right. \nonumber
\end{align}

\subsubsection{OP-MM System}
It can be shown that the harvested energy by user $k$ is $E_{k,\text{OP-MM}} (\alpha) = \alpha p_{dl} \beta_k$. By replacing $E_k(\alpha, \xi_k, \rho)$ in~\eqref{eq:RateLBZF} by $E_k(\alpha)$, the achievable rate for ZF is given by
\begin{align}
  \tilR_{k,\text{OP-MM}}^{\text{ZF}}  (\tau, \alpha, \rho) &= (1 \!-\! \tau \!-\! \alpha) \log \left( 1 + \tilgamma_{k,\text{OP-MM}}^{\text{ZF}} \right),  \label{eq:RateLBZFRandomBF} 
\end{align}
where the SINR
\[\tilgamma_{k,\text{OP-MM}}^{\text{ZF}} = \frac{(M-K) \alpha p_{dl} \beta_k^3 \rho}{\sigma^2 \big(\beta_k \rho +\frac{\sigma^2}{\alpha p_{dl} \beta_k}\big) \left( \frac{1-\tau-\alpha}{1-\rho} + \sum \limits_{i=1}^K \frac{\alpha p_{dl} \beta_i^2}{\alpha p_{dl} \beta_i^2 \rho +\sigma^2} \right)}.
\]

Similarly, from~\eqref{eq:RateLBMRC}, the achievable rate for MRC is thus given by
\begin{align}
  \tilR_{k,\text{OP-MM}}^{\text{MRC}} (\tau, \alpha, \rho) &=(1 \!-\! \tau \!-\! \alpha) \log \left( 1 + \tilgamma_{k,\text{OP-MM}}^{\text{MRC}} \right), \label{eq:RateLBMRCRandomBF} 
\end{align}
where the SINR
\begin{align}
&\tilgamma_{k,\text{OP-MM}}^{\text{MRC}} = \nonumber \\
&\frac{(M-1) \alpha p_{dl} \beta_k^3 \rho }{\big(\beta_k \rho +\frac{\sigma^2}{\alpha p_{dl} \beta_k}\big) \left( \frac{\sigma^2 (1-\tau-\alpha)}{1-\rho}+ \sum \limits_{i=1, i \neq k}^K \alpha p_{dl} \beta_i^2 \right) + \beta_k \sigma^2}. \nonumber
\end{align}

\section{Asymptotic Analysis}\label{SecAsymAnalysis}
In this section, we analyze the UL achievable rate in the large-$M$ regime.

\subsection{Asymptotic Analysis for WET-MM System with Estimated CSI} \label{Sec:AsympAna}

To obtain analytical insights, we consider the massive MIMO regime where the number of transmit antennas at H-AP, $M$, is sufficiently large. Specifically, we assume
\begin{align}
  M \gg \underset{1 \leq k \leq K}{\max} \; \frac{\sigma^2}{\alpha p_{dl} \rho \beta_k^2 \xi_k}. \label{eq:conditionM}
\end{align}
Then, from~\eqref{HarvestedEnergy1}, the asymptotic harvested energy at user $k$ is given by
\begin{align}
  E_k(\alpha, \xi_k, \rho) \convergeM E_k^{\textrm{asym}}(\alpha, \xi_k) \triangleq \alpha p_{dl} \beta_k \xi_k M. \label{eq:EnergyAsym}
\end{align}

\begin{myrem}[Discussion on the asymptotically harvested energy]\label{RemarkAsympEnergy}
  The asymptotically harvested energy $E_k^{\textrm{asym}}$ is achieved when $M$ is sufficiently large such than the $M$-dependent term in~\eqref{HarvestedEnergy1} is dominant over other terms. We note that the $E_k^{\textrm{asym}}$ is independent of $\rho$, as the $\rho$-dependent terms in~\eqref{HarvestedEnergy1} are negligible.

  For the condition~\eqref{eq:conditionM}, we assume that $\alpha, \; \rho$ and $\xi_k$ are arbitrarily fixed and independent\footnote{Later we shall consider $\alpha, \rho, \xi_k$ as variables to be optimized.} of $M$. However, we noted that when $\alpha = O (M^{-2 \nu})$ for $0 < \nu < \frac{1}{2}$, the condition~\eqref{eq:conditionM} still holds, and $E_k^{\textrm{asym}}(\alpha, \xi_k)=O(M^{1-2\nu})$ increases as $M$ increases. This observation will be used in Section~\ref{SecAsymOptSolution}.

Also, we observe that the asymptotically harvested energy in~\eqref{eq:EnergyAsym} approaches the harvested energy in~\eqref{eq:EnergyIdealCase} for the ideal case with perfect CSI. That is, with the proposed scheme of energy splitting between UL CE and UL WIT, the energy beamforming with estimated CSI asymptotically achieves the DL-WET performance limit that is achieved by the ideal case.
\end{myrem}

Moreover, the asymptotic rate is given in Theorem~\ref{TheoremZFAsymp} for ZF, and in  Theorem~\ref{TheoremMRCAsymp} for MRC.

\begin{mythe}\label{TheoremZFAsymp}
For fixed $\tau, \; \alpha, \;\rho$ and $K$, when $M$ is sufficiently large such that~\eqref{eq:conditionM} is satisfied, the asymptotically achievable rate of user $k$ for ZF detection is given by
\begin{align}
    \tilR_k^{\sf ZF} &\convergeM  (1-\tau-\alpha) \log \left( 1+ \frac{M (M-K) \alpha p_{dl} \beta_k^2 \xi_k \rho}{ \sigma^2 \big[ K+ \frac{(1-\tau-\alpha) \rho}{1-\rho} \big]} \right). \label{eq:RateLBZFAsymp1}
  \end{align}
\end{mythe}

\begin{IEEEproof}
Taking $M \rightarrow \infty$ and assuming~\eqref{eq:conditionM} holds, from~\eqref{eq:RateLBZF} in Lemma~\ref{LemmaZF} and~\eqref{eq:EnergyAsym}, the asymptotically achievable rate is derived in~\eqref{eq:RateLBZFAsymp1}.
\end{IEEEproof}


\begin{mythe}\label{TheoremMRCAsymp}
For fixed $\tau, \; \alpha, \;\rho$ and $K$, when $M$ is sufficiently large such that
\begin{align}
  M \gg \; \underset{1 \leq k \leq K}{\max} \; &\max \; \Bigg\{ \frac{\sigma^2}{\alpha p_{dl} \rho \beta_k^2 \xi_k}, \; \frac{\sigma^2 (1-\tau-\alpha)}{\alpha p_{dl} \beta_k \xi_k (1-\rho)} \Bigg\}, \label{eq:conditionMMRC}
\end{align}
the asymptotically achievable rate  of user $k$ for MRC detection is given by
\begin{align}
    \tilR_k^{\sf MRC} &\convergeM  (1 \!-\! \tau \!-\! \alpha) \log \left( 1+ \frac{(M-1) \beta_k^2 \xi_k}{\sum \limits_{i \neq k} \beta_i^2 \xi_i} \right). \label{eq:RateLBMRCAsymp1}
  \end{align}
\end{mythe}

\begin{IEEEproof}
Taking $M \rightarrow \infty$, and assuming $M \gg \underset{1 \leq k \leq K}{\max} \; \frac{\sigma^2 (1-\tau-\alpha)}{\alpha p_{dl} \beta_k \xi_k (1-\rho)}$ and~\eqref{eq:conditionM} holds, from~\eqref{eq:EnergyAsym}, the noise power in~\eqref{eq:RateLBMRC} in Lemma~\ref{LemmaZF} is negligible compared to the MUI power, and the asymptotically achievable rate is further derived in~\eqref{eq:RateLBMRCAsymp1}. 
\end{IEEEproof}
For MRC detection, the asymptotic rate is independent of the energy-splitting fraction $\rho$. This is because both the signal power and the MUI power in~\eqref{eq:RateLBMRC} are proportional to $\rho$. The effect of $\rho$ is thus cancelled. Thus, $\rho$ can be arbitrarily chosen in $(0,1)$. Also, we note that the condition in~\eqref{eq:conditionMMRC} can be easily satisfied as long as $\rho$ does not approach 1.


For the proposed WET-MM system, the asymptotic SINR in~\eqref{eq:RateLBZFAsymp1} for ZF is of order $O \left( M^2 \right)$, since the MUI is cancelled. For MRC, however,  the asymptotic SINR in~\eqref{eq:RateLBMRCAsymp1} is of order $O \left( M \right)$, due to the MUI. The factor $(M-1)$ in the asymptotical SINR is due to the maximum ratio combining. The MM-DoDR of the proposed WET-MM system is thus given immediately by the following theorem.
\begin{mythe}\label{TheoremZFMRCDoRG}
  For a WET-MM system, the maximal asymptotic MM-DoDR of any user $k$ is given by
\begin{align}
  \kappa_{\sf WET-MM} = \left\{
  \begin{array}{cl}
    2, \quad &\mbox{for ZF} \\
    1, \quad &\mbox{for MRC}.
  \end{array}
  \right.\label{eq:MMDoRGWETMM}
\end{align}
\end{mythe}
\begin{IEEEproof}
  Clearly, from Theorem~\ref{TheoremZFAsymp} and Theorem~\ref{TheoremMRCAsymp}, we have
  \begin{align}
  \kappa_{\text{WET-MM}} = \lim \limits_{M \rightarrow \infty} \frac{\tilR_k}{ \log{M} } = \left\{
  \begin{array}{cl}
    2 (1-\tau-\alpha), \quad &\mbox{for ZF} \\
    1-\tau-\alpha, \quad &\mbox{for MRC}. \nonumber
  \end{array}
  \right.
\end{align}
For ZF, let $\tau \rightarrow 0$, \ $\alpha=O \left(M^{-2\nu}\right) \rightarrow 0$, where $0 < \nu < \frac{1}{2}$. For MRC, let $\tau \rightarrow 0$, and $\alpha=O \left(M^{-\varphi}\right) \rightarrow 0$, where $0 < \varphi < 1$. The maximal asymptotic MM-DoRG is thus obtained in~\eqref{eq:MMDoRGWETMM}.
\end{IEEEproof}

\subsection{Asymptotic Analysis for Ideal Case and OP-MM System} \label{Sec:AsympAnaBenchmarks}
The MM-DoRG for the ideal case and the OP-MM system is given in the following theorem.
\begin{mythe}\label{TheIdealCase_DoRG}
The maximal asymptotic MM-DoDR for the OP-MM system is given by
\begin{align}
  \kappa_{\sf OP-MM}  = 1, \quad &\mbox{for ZF and MRC}. \label{eq:DoRGOT}
\end{align}
The maximal asymptotic MM-DoDR for the ideal case of the WET-MM system is given by
\begin{align}
  \kappa_{\sf {Ideal}}  = \left\{
  \begin{array}{cl}
    2, \quad &\text{for ZF} \\
    1, \quad &\text{for MRC}. \\
  \end{array}
  \right. \label{eq:DoRGPCSI}
\end{align}
\end{mythe}
\begin{IEEEproof}
For the OP-MM system, the asymptotic achievable rate is still given in~\eqref{eq:RateLBZFRandomBF} for ZF, and in~\eqref{eq:RateLBMRCRandomBF} for MRC. Note that the asymptotic SINR is of order $O \left( M \right)$. By definition, the maximal asymptotic MM-DoDR for the OP-MM system is obtained in~\eqref{eq:DoRGOT}.

For the ideal case of the WET-MM system, from~\eqref{eq:EnergyIdealCase}, we obtain that the harvested energy $E_k(\alpha, \xi_k) \convergeM \alpha p_{dl} \xi_k \beta_k M$.
Then, the achievable rate for ZF is given from~\eqref{eq:RateLBMRCZFPerfectCSI0} by
\begin{align}
  \tilR_k^{\text{ZF}} (\alpha, \xi_k) &\convergeM (1-\alpha) \log \left( 1+ \frac{\alpha p_{dl} \beta_k^2 \xi_k M (M-K) }{\sigma^2 (1-\alpha)} \right).  \label{eq:RateLBZFPerfectCSI1}
\end{align}
The achievable rate for MRC is given from~\eqref{eq:RateLBMRCZFPerfectCSI0} by
\begin{align}
  \tilR_k^{\text{MRC}} (\alpha, \bm{\xi}) &\convergeM (1 \!-\! \alpha) \log \left(\!  1 \!+\! \frac{(M \! - \! 1) \beta_k^2 \xi_k}{\sum \limits_{i=1, i \neq k}^K \beta_i^2 \xi_i} \! \right). \label{eq:RateLBMRCPerfectCSI2}
\end{align}
The asymptotic SINR of ZF in~\eqref{eq:RateLBZFPerfectCSI1} is of order $O \left( M^2 \right)$, while the asymptotic SINR for MRC in~\eqref{eq:RateLBMRCPerfectCSI2} is of order $O \left( M \right)$. By definition, the maximal asymptotic MM-DoDR is thus obtained in~\eqref{eq:DoRGPCSI}.
\end{IEEEproof}
\begin{myrem}[Advantages over the OP-MM system] The proposed WET-MM system outperforms the OP-MM system in the following three aspects. First, the MM-DoRG achieved by the WET-MM system $\kappa_{\sf{WET-MM}}$ is double of that achieved by the OP-MM system $\kappa_{\sf{OP-MM}}$. This is because more harvested energy (due to energy beamforming) leads to both more accurate CSI and increased uplink transmit power in the WIT phase. From Theorem~\ref{TheoremZFMRCDoRG} and Theorem~\ref{TheIdealCase_DoRG}, we can interpret the MM-DoRG metric as the proportional gain of the asymptotic rate for a massive MIMO system power by WET with beamforming, with respect to the OP-MM system. Second, we shall see in Remark~\ref{RemarkFairnessComparison} in Section~\ref{SecAsymOptSolution} that the proposed WET-MM system asymptotically achieves the best possible rate fairness among users (i.e., a common rate), while no fairness is guaranteed in the OP-MM system. Third, as will be numerically shown in Section~\ref{Simulation}, to achieve a desired common rate for all users with a given maximal AP-user distance, the proposed WET-MM system requires less antennas at the H-AP, roughly a square root of that required by the OP-MM system.
\end{myrem}

\section{Asymptotically Optimal Solution}\label{SecAsymOptSolution}
Following the asymptotic analysis in Section~\ref{SecAsymAnalysis}, in this section, we derive the asymptotically optimal solution to the minimum rate maximization Problem $(\mathrm{P_1})$ in Section~\ref{Sec:SolutionProposed} for the proposed WET-MM system, and in Section~\ref{Sec:SolutionBenchmark} for the ideal case and the OP-MM system.
\subsection{Asymptotically Optimal Solution for WET-MM Systems}\label{Sec:SolutionProposed}

\subsubsection{Asymptotically Optimal Energy Allocation Weights} \label{SecOptXiDerive}
In the large-$M$ regime, the optimal $\xi_k$ that maximizes the minimum rate is given in Lemma~\ref{Lemma:CoeffDesign}.
\begin{mylem}\label{Lemma:CoeffDesign}
When $M$ satisfies the condition in~\eqref{eq:conditionM}, for both ZF and MRC detection, the minimum achievable rate is maximized when the energy allocation weight is chosen as
\begin{align}
      \xi_k^{\star} = \frac{\frac{1}{\beta_k^2}}{\sum \limits_{i=1}^K \frac{1}{\beta_i^2}}. \label{eq:coefficientXi}
\end{align}
\end{mylem}

\begin{IEEEproof}
  See Appendix~\ref{AppendixCoeff}.
\end{IEEEproof}
That is, the optimal $\xi_k^{\star}$ is inversely proportional to the square of the long-term path loss of user $k$, which compensates the long-term path loss in both the downlink WET phase and the uplink WIT phase. This optimal $\xi_k^{\star}$ enables users to asymptotically achieve a common rate (see Theorem~\ref{Theorem:OptRate} later), thus ensuring fairness among users.

Substituting~\eqref{eq:coefficientXi} into~\eqref{eq:conditionM}, we observe that $M$ should be much greater than some
constant, given $\alpha$ and $\rho$. As numerically shown in Section~\ref{Simulation}, for our analysis to hold, the condition in~\eqref{eq:conditionM} is well satisfied for $M$ as small as 25.
\subsubsection{Asymptotically Optimal Energy-Splitting}\label{eq:OptRhoMaxmin}
Here, we assume $\tau$ and $\alpha$ to be arbitrary. We first obtain the asymptotically optimal $\rho^{\star}$ for Problem ($\mathrm{P_1}$) in the following lemma.
\begin{mylem}\label{Lemma:OptRho}
For ZF detection, when $M$ satisfies~\eqref{eq:conditionM}, the asymptotically optimal $\rho_{\text{ZF}}^{\star}$ is given by 
\begin{align}
      \rho_{\text{ZF}}^{\star}(\tau, \alpha) =
      \frac{\sqrt{K}}{\sqrt{K}+\sqrt{1-\tau-\alpha}}.
\label{eq:OptRhoZF}
\end{align}
For MRC detection, when $M$ satisfies~\eqref{eq:conditionMMRC}, the asymptotically optimal energy-splitting fraction $\rho_{\text{MRC}}^{\star}$ that maximizes the achievable rate is arbitrary in $(0,1)$.
\end{mylem}

\begin{IEEEproof}
For ZF detection, given $\tau, \; \alpha$, the asymptotic SINR in~\eqref{eq:RateLBZFAsymp1} is proportional to the function 
\[
f(\rho)= \frac{\rho}{ K+ \frac{(1-\tau-\alpha) \rho}{1-\rho}}.
\]
It can be shown that $f(\rho)$ is a quasiconcave function of $\rho$. The optimal $\rho^{\star}$ that maximizes $f(\rho)$ is obtained by solving $f'(\rho)=0$ where $f'(\rho)$ is the derivative of $f(\rho)$. The solution is given by~\eqref{eq:OptRhoZF}.%

For MRC detection, the asymptotic rate in~\eqref{eq:RateLBMRCAsymp1} is independent of $\rho$. Hence, the energy-splitting coefficient $\rho$ can be arbitrary chosen in $(0,1)$.
\end{IEEEproof}

For ZF detection, we observe that the asymptotically optimal $\rho_{\text{ZF}}^{\star}$ increases as the number of users $K$ increases. It implies that a higher fraction of harvested energy should be used for CE for larger $K$. This is because the asymptotic rate is more interference-limited for large $K$, and hence more accurate CSI is required to decrease the interference that comes from channel estimation error.

\subsubsection{Asymptotically Optimal Time Allocation}
The asymptotically optimal $\tau^{\star}$ and $\alpha^{\star}$ is given in the following Lemma~\ref{OpttauAlphaZFAsymp} for ZF detection, and in Lemma~\ref{OpttauAlphaMRCAsymp} for MRC detection.
\begin{mylem}\label{OpttauAlphaZFAsymp}
For ZF detection, when $M$ is sufficiently large to satisfy~\eqref{eq:conditionM}, the asymptotically optimal time allocation for CE and WET is given by
\begin{align}
  \tau_{\text{ZF}}^{\star} \convergeM 0, \quad
&\alpha_{\text{ZF}}^{\star} = O\left({M^{-2 \nu}}\right) \convergeM 0, \; \nonumber \\ &\text{where}\; \nu > 0, \; \text{and} \; \nu \rightarrow 0. \label{eq:AsymZFtauAlpha}  
\end{align}
\end{mylem}

\begin{IEEEproof}
Suppose the condition~\eqref{eq:conditionM} is satisfied. We define the asymptotic SINR in~\eqref{eq:RateLBZFAsymp1} as a function $g(\tau)$ of $\tau$. Then the derivative of $g(\tau)$ is derived as
  \begin{align}
    g'(\tau) &= \frac{(1-\tau-\alpha)\rho}{ {(1-\rho) K} + (1-\tau-\alpha) \rho } - \nonumber \\
    &\quad \quad \log \left( \frac{M(M-K)\alpha p_{dl} \beta_k^2 \xi_k \rho}{\sigma^2 \big[ K + \frac{(1-\tau-\alpha) \rho}{1-\rho} \big]} \right). \nonumber
  \end{align}
Clearly, the above derivative is negative, due to the assumption of large $M$. Hence, we choose $\tau_{\text{ZF}}^{\star} \rightarrow 0$ to maximize $g(\tau)$.

Suppose $\alpha = O(M^{-2\nu})$, for $0 < \nu < \frac{1}{2}$; otherwise, the asymptotically harvested energy in~\eqref{eq:EnergyAsym} decreases as $M$ increases, which is suboptimal. Define $\epsilon(M)$ such that $\epsilon(M) \rightarrow 0$ for sufficiently large $M$. From~\eqref{eq:RateLBZFAsymp1}, the asymptotic rate is rewritten as
\begin{align}
    \tilR_k^{\text{ZF}} &\convergeM  \big(\!1 \!-\! \tau \!-\! \epsilon(M)\big) \log \left(\! \! 1 \!+\! \frac{M^{2(1-\nu)} p_{dl} \beta_k^2 \xi_k \rho}{ \sigma^2 \big[ K \!+ \! \frac{(1-\tau-\epsilon(M)) \rho}{1-\rho} \big]} \! \right) \nonumber \\
    & \convergeM (1-\tau) \log \left( 1+ \frac{M^{2(1-\nu)} p_{dl} \beta_k^2 \xi_k \rho}{ \sigma^2 \big[ K+ \frac{ (1-\tau) \rho}{1-\rho} \big]} \right).
     \label{eq:RateLBZFAsymp2}
  \end{align}
We choose $\nu \rightarrow 0$ to maximize the asymptotic rate. Hence, $\alpha_{\text{ZF}}^{\star} = O(M^{-2\nu})$, where $\nu \rightarrow 0$.
\end{IEEEproof}

\begin{mylem}\label{OpttauAlphaMRCAsymp}
For MRC detection, when $M$ is sufficiently large to satisfy~\eqref{eq:conditionMMRC}, the asymptotically optimal time allocation for CE and WET is given by
\begin{align}
  \tau_{\text{MRC}}^{\star} \convergeM 0, \quad
\alpha_{\text{MRC}}^{\star} = O\left({M^{-\varphi}}\right) \rightarrow 0, \label{eq:AsymMRCtauAlpha}
\end{align}
where $0 < \varphi < 1$ is arbitrary.
\end{mylem}

\begin{IEEEproof}
From~\eqref{eq:RateLBMRCAsymp1}, the asymptotically optimal $\tau$ and $\alpha$ should be chosen as small as possible. To satisfy~\eqref{eq:conditionMMRC}, we choose $\alpha$ to tend to zero at the rate of order of $O\left(M^{-\varphi}\right)$, for $0 < \varphi < 1$.
\end{IEEEproof}

For ZF detection, the optimal WET time $\alpha_{\text{ZF}}^{\star} $ in Lemma~\ref{OpttauAlphaZFAsymp} approaches zero fairly slowly with the number of antennas $M$, as will be verified numerically in Section~\ref{Simulation}. For MRC detection, however, the optimal $\alpha_{\text{MRC}}^{\star} $ in Lemma~\ref{OpttauAlphaMRCAsymp} approaches zero faster with $M$, when $\varphi$ is chosen to approach one.

\vspace{0.4cm}
\subsubsection{Asymptotically Maximal Minimum Rate}
For both ZF detection and MRC detection, the asymptotically maximal minimum achievable rate for Problem ($\mathrm{P_1}$) is thus given in the following theorem.
\begin{mythe}\label{Theorem:OptRate}
The asymptotically maximal minimum rate for the proposed WET-MM system is
\begin{align}
    \tilR \convergeM \left\{ \begin{array}{cl}
    \log \left( 1+ \frac{M^{2} p_{dl}}{ \sigma^2 ( \sqrt{K}+ 1)^2 \sum \nolimits_{i=1}^K \frac{1}{\beta_i^2}} \right),  &\mbox{for ZF} \\
     \log \Big( 1+ \frac{M-1}{K-1} \Big), &\mbox{for MRC.}
\label{eq:MaxMinRate}
    \end{array}
    \right.
  \end{align}
\end{mythe}

\begin{IEEEproof}
For MRC, from Lemma~\ref{OpttauAlphaMRCAsymp} and Theorem~\ref{TheoremMRCAsymp}, the asymptotically maximal rate for user $k$ is obtained in~\eqref{eq:MaxMinRate}. For ZF, from Lemma~\ref{Lemma:OptRho} and Lemma~\ref{OpttauAlphaZFAsymp}, the asymptotically optimal $\rho_{\text{ZF}}^{\star}= \frac{\sqrt{K}}{\sqrt{K}+1}$. From Lemma~\ref{OpttauAlphaZFAsymp} and Theorem~\ref{TheoremZFAsymp}, the asymptotically maximal rate for user $k$ is obtained in~\eqref{eq:MaxMinRate}. The fact that each user achieves the same asymptotically maximal rate completes the proof.
\end{IEEEproof}


\begin{myrem}[Fairness comparison] \label{RemarkFairnessComparison}
The proposed WET-MM system asymptotically achieves the best possible rate fairness among users (i.e., a common rate), while no fairness is guaranteed in the OP-MM system. For the OP-MM system, from~\eqref{eq:RateLBZFRandomBF}, the rate achieved by users located far away may be extremely low, since few energy is harvested.
\end{myrem}


\begin{myrem}[Asymptotic performance for large $M$ and $K$]\label{RemarkLargeK}
Previously, the number of users $K$ is fixed. Herein, we let $K$ scale up with $M$ while keeping a constant ratio $\frac{K}{M} \triangleq \zeta \in(0,1)$. As $M, K \rightarrow \infty$ and $\frac{K}{M}=\zeta$, similarly, we obtain the asymptotic rate as follows
\begin{align} 
    \tilR \convergeM
    &\log \left( \frac{\alpha^{\star} p_{\sf dl} (1-\zeta)}{c_1  \sigma^2 \zeta^2}\right) \triangleq \tilR ({\zeta}), \label{eq:AsymRate_Scale_K}
  \end{align}
where $c_1=\frac{1}{K} \sum \nolimits_{i=1}^K \frac{1}{\beta_i^2}$. We assume the path loss is modeled as $\beta_k = \beta_0 d_k^{-u}$, where $\beta_0$ is the loss for a distance of one meter, and $u > 1$ is the path-loss exponent. We then have the following claim on the quantity $c_1$.

\noindent {{\textit{Claim 1. }}}Assuming the distance $d_k$ is independent and uniformly distributed in the interval $[a, b]$, where $0< a < b$, it holds that $c_1$ approaches a constant for large $K$, i.e., $c_1 \rightarrow \frac{\beta_0^{-2} \left( b^{2u+1} - a^{2u+1} \right)}{(b-a)(2u+1)}$, as $K \rightarrow \infty$.

\begin{IEEEproof}
From the path loss model, we have that $c_1= \frac{\beta_0^{-2}}{K} \sum \nolimits_{i=1}^K Y_i$, where $Y_i = d_i^{2u}$. The probability density function of random variable $Y_k$ is obtained as $f_Y(y)=\frac{y^{(0.5-u)/u}}{2u(b-a)}$, for $a^{2u} \leq y \leq b^{2u}$, and $f_Y(y)=0$ otherwise. It is standard to show that $\bbE [Y_i]=\frac{(b^{2u+1}-a^{2u+1})}{(2u+1)(b-a)}$. The desired result is obtained by using the independence among $Y_i$'s and the law of large numbers.
\end{IEEEproof}

In the regime of large $M$ and $K$, for a desired common rate $R_0$, the asymptotically supporting number of users $K$ is given by $\zeta_0$, where $\zeta_0$ is the solution to $\tilR ({\zeta})=R_0$ for $\zeta$. It can be shown that the positive solution $\zeta_0$ is unique.

In the ideal case, the H-AP has perfect CSI, without requiring users to send pilots. Comparatively, the proposed WET-MM system with larger $K$ suffers inevitable performance degradation, since users have to spend more time and energy for sending pilots (as implied in Lemma~\ref{Lemma:OptRho}). However, the proposed WET-MM system achieves the best achievable performance in the present optimization framework with imperfect CSI, in which we vary the allocation of time and energy resources, and the fraction of harvested energy used for sending pilots.


\end{myrem}

\subsection{Asymptotically Optimal Solutions for Ideal Case and OP-MM System} \label{Sec:SolutionBenchmark}
\subsubsection{Ideal Case}
Similar to Section~\ref{SecOptXiDerive}, it can be shown that the asymptotically optimal energy allocation weights $\bm{\xi}^{\star}$ is given by~\eqref{eq:coefficientXi}.
Thus, from~\eqref{eq:RateLBZFPerfectCSI1} and~\eqref{eq:RateLBMRCPerfectCSI2}, the achievable rate is given by
\begin{align}
  &\tilR_{k,\text{Ideal}} (\alpha) \convergeM \nonumber \\
  &\; \left\{
  \begin{array}{cl}
              (1 \!-\! \alpha) \log \left( 1 \!+\! \frac{\alpha p_{dl} M (M-K) }{\sigma^2 (1-\alpha) \sum \limits_{i=1}^K \frac{1}{\beta_i^2}} \right), \! &\mbox{for ZF} \\
       (1 \!-\! \alpha) \log \left( 1 \!+\! \frac{M-1}{K-1} \right), \; &\mbox{for MRC.} \label{eq:RateLBZFMRCPerfectCSI1}
  \end{array}
  \right. %
\end{align}
For ZF detection, similar to Lemma~\ref{OpttauAlphaZFAsymp}, it can be shown from~\eqref{eq:RateLBZFMRCPerfectCSI1}, the optimal $\alpha_{\text{Ideal}}^{\star} = O \left( {M^{-2 \nu}} \right) \rightarrow 0$, where $\nu > 0$, and $\nu \rightarrow 0$.
For MRC, similar to Lemma~\ref{OpttauAlphaMRCAsymp}, from~\eqref{eq:RateLBZFMRCPerfectCSI1}, we have that the optimal $\alpha_{\text{Ideal}}^{\star} = O\left({M^{-\nu}}\right) \rightarrow 0$, where $0 < \nu < 1$ is arbitrary.

\subsubsection{OP-MM System}
For both ZF and MRC detection, the achievable rates in~\eqref{eq:RateLBZFRandomBF} and~\eqref{eq:RateLBMRCRandomBF} are functions of $\tau, \alpha$ and $\rho$. Hence, the asymptotically optimal $\tau_{\text{OT}}^{\star}, \alpha_{\text{OT}}^{\star}$ and $\rho_{\text{OT}}^{\star}$ that maximize the achievable rate can be obtained by a numerical search.


\section{Numerical Results}\label{Simulation}
In this section, we present numerical results to validate our results. We employ ZF detection given in this paper, as well as MRC detection, see details in~\cite{YangHoGuanMMIMO14}. We set $K=2$ and $p_{\sf dl}=1 \ \text{Watt}$. The frame length is fixed as $5$ $\mu$s, which is normalized in the simulations. The carrier frequency is $5$ GHz, and the bandwidth is $100$ KHz. We set the power spectrum density of noise as $-170 \ \textrm{dBm/Hz}$, which implies the noise power $\sigma^2 = -120 \ \textrm{dBm}$. We use the long-term fading model $\beta_i = 10^{-3} d_i^{-3}$, where the distance $d_1=6$ m and $d_2=12$ m. We use 1000 channel realizations in the Monte Carlo simulation. The step size for $\tau$, $\alpha$ and $\rho$ is chosen as $0.00025, 0.0005$ and $0.0005$, respectively.

\begin{figure} 
\centering
\includegraphics[width=1.07\columnwidth] {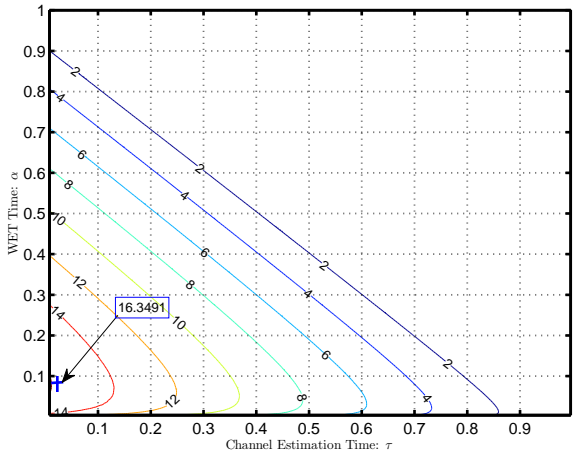}
\caption{Rate of user 1 versus CE time $\tau$ and WET time $\alpha$.}
\label{fig:Fig2a}
\end{figure}

\begin{figure}
\centering
\includegraphics[width=1.07\columnwidth] {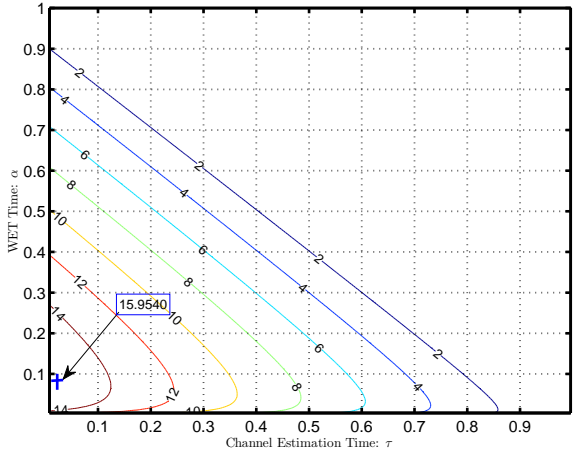}
\caption{Rate of user 2 versus CE time $\tau$ and WET time $\alpha$.}
\label{fig:Fig2b}
\end{figure}

\begin{figure} 
\centering
\includegraphics[width=1.07\columnwidth] {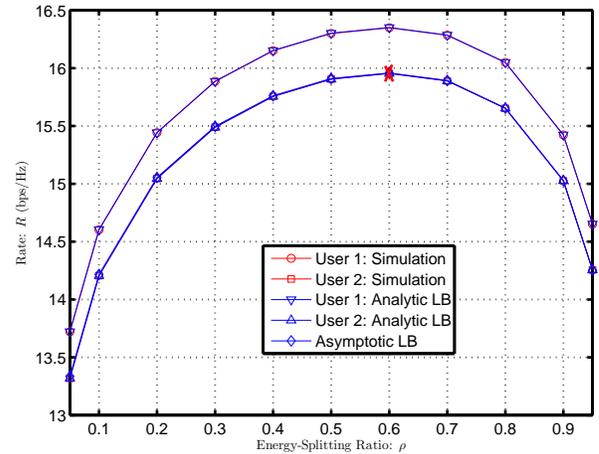}
\caption{Effect of energy-splitting fraction $\rho$ on data rate.}
\label{fig:Fig2aa}
\end{figure}

First, we verify the asymptotically optimal solution for ZF detection. We fix $M=200$. By numerical search, the optimal time for CE and for WET is $\tau^{\star}=0.00825, \alpha^{\star}=0.0760$, respectively. From the analytic result~\eqref{eq:OptRhoZF} in Lemma~\ref{Lemma:OptRho}, the asymptotically optimal $\rho^{\star}=0.5961$. From~\eqref{eq:MaxMinRate} in Theorem~\ref{Theorem:OptRate}, the maximal minimum rate is thus $16.0096$ bps/Hz. These results will be verified in Figs.~\ref{fig:Fig2a}, Fig.~\ref{fig:Fig2b} and Fig.~\ref{fig:Fig2aa}.

For fixed $\rho=0.5965$, Fig.~\ref{fig:Fig2a} and Fig.~\ref{fig:Fig2b} are the contour plot of the rate against the uplink CE time $\tau$ and the downlink WET time $\alpha$, for user 1 and 2, respectively. Specifically, the maximal minimum rate is achieved as $15.9540$ bps/Hz, at $\tau^{\star}=0.00825, \alpha^{\star}=0.0760$. Furthermore, the achieved rate is $16.3491$ bps/Hz and $15.9540$ bps/Hz for user 1 and user 2, respectively, which is almost the same. The asymptotically maximal minimum rate is $16.0096$ bps/Hz, which approximates the actually maximal minimum rate $15.9540$ bps/Hz closely. Hence, the asymptotic result in Theorem~\ref{Theorem:OptRate} is numerically verified.

For fixed time allocation $\tau^{\star}=0.00825, \alpha^{\star}=0.0760$, the data rate is plotted in Fig.~\ref{fig:Fig2aa} against the energy-splitting fraction $\rho$. We see that the rate is a quasi-concave function of $\rho$, and the unique optimal energy-splitting fraction $\rho^{\star} \approx 0.6$ is the star-marker point in Fig.~\ref{fig:Fig2aa}. This coincides with the analytic result in Lemma~\ref{Lemma:OptRho}.

Moreover, we compare the optimal resource allocation solutions for different $M$. The numerical results are shown in Table~\ref{table1}. We use the normal notation for the analytically asymptotic results, and use the notations with the bar head for the numerical results. We observe that as $M$ tends to infinity, the optimal CE time $\bar{\tau}^{\star}$ tends to 0 quickly, while the optimal WET time $\bar{\alpha}^{\star}$ tends to 0 at a slower rate. This observation coincides with the results in Lemma~\ref{OpttauAlphaZFAsymp}. From Lemma~\ref{Lemma:OptRho} and Lemma~\ref{OpttauAlphaZFAsymp}, we obtain that the asymptotically optimal energy-splitting ratio $\rho^{\star}=0.5961$ for $M=200$, which is approached by the numerically obtained optimal $\bar{\rho}^{\star}$ in~Table~\ref{table1}.
\begin{table*} [htbp]
\centering
\caption{Optimal resource allocation versus $M$} \label{table1}
\small{
\begin{tabular}{*{9}{c}}
  \hline \hline
  $M$ & $\bar{\tau}^{\star}$ & $\bar{\alpha}^{\star}$  & $\rho^{\star}$ & $\bar{\rho}^{\star}$ &  $\tilR_{\text{asym.}}^{\star}$  &  $\tilR_{\text{anal.}}^{\star}$ & $\bar{R_1}$ & $\bar{R_2}$ \\
  \hline
  $25$   & 0.03750 & 0.1455  & 0.6125 & 0.6425  & 10.4597  & 10.5162 & 12.1786 & 10.5635\\
  $50$   & 0.02725 & 0.1180  & 0.6075 & 0.6250  & 12.3200  & 12.3327 & 13.4457 & 12.3319\\
  $100$  & 0.01875  & 0.0945  & 0.6025 & 0.6025  & 14.1650  & 14.2022 & 14.7690 & 14.2114\\
  $200$  & 0.00825  & 0.0760 & 0.5961 & 0.5965 & 16.0096  & 16.0098 & 16.3491 & 15.9540 \\
  $400$  & 0.00475 & 0.0580 & 0.5943 & 0.5950 & 17.8603 & 17.8677 & 18.0551 & 17.8744 \\
  $600$  & 0.00275  & 0.0515 & 0.5936 & 0.5930 & 18.9469 & 18.9490 & 19.0955 & 18.9520 \\
  $800$  & 0.00125 & 0.0505 & 0.5912 & 0.5915 & 19.7195 & 19.7147 & 19.8604 & 19.7240 \\
  $1000$ & 0.00075  & 0.0490 & 0.5901  & 0.5905 & 20.3196 & 20.3227 & 20.4010 & 20.3233\\
  \hline
\end{tabular}
}
\end{table*}

\begin{figure}
\centering
\includegraphics[width=1.07\columnwidth] {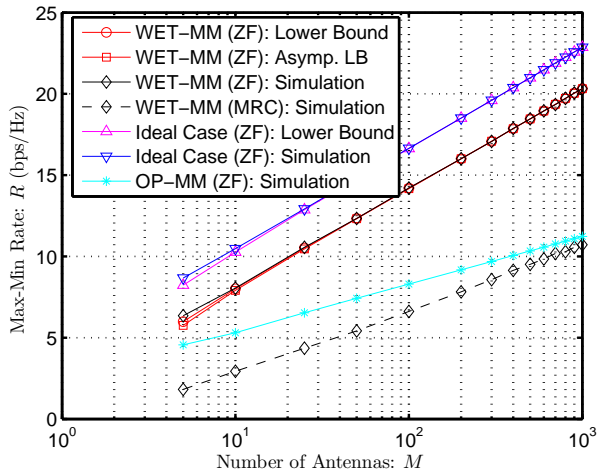}
\caption{Comparison of achievable rates versus $M$.}
\label{fig:Fig4}
\end{figure}

Next, we simulate the achievable rates for ZF detection and MRC detection.
Fig.~\ref{fig:Fig4} compares the achievable rate to that of the ideal case and of the OP-MM system, where the x-axis is in the logarithm scale. For ZF detection, we observe that the proposed WET-MM system achieves more than $80\%$ of the rate limit achieved by the ideal case with perfect CSI. Also, we see that the proposed WET-MM system achieves higher rates than the OP-MM system. These observations numerically show the efficiency of the proposed system. Moreover, we observe that the analytically-obtained rate approaches that obtained from simulation, even when $M$ is as small as 10. The asymptotic analysis is thus accurate even for the case of small $M$. Finally, we see that the achievable rate for ZF detection is higher than that for MRC detection, which is as expected.

Recall that the MM-DoRG $\kappa$ is defined in~\eqref{eq:MMDoRG} as the asymptotic slope of rate $R$ with respect to $\log{M}$. We then observe from Fig.~\ref{fig:Fig4} that the MM-DoRG of the proposed WET-MM system achieves the same MM-DoRG as the ideal case, i.e., $\kappa_{\sf{WET-MM}}=\kappa_{\sf{Ideal}}=2$, which is double of the MM-DoRG of the OP-MM system. Hence, the proposed WET-MM system is numerically verified to be optimal in terms of MM-DoRG.

Also, we observe from Fig.~\ref{fig:Fig4} that to achieve a max-min rate at $10, 8, 6.4$ bps/Hz for the maximal distance $d_2=12$ m, the OP-MM system requires about 400, 100, 25 antennas at the H-AP, while the proposed WET-MM requires only 21, 10, 5 antennas, respectively, being roughly a square root of that required by OP-MM system.

\begin{figure}
\centering
\includegraphics[width=1.07\columnwidth] {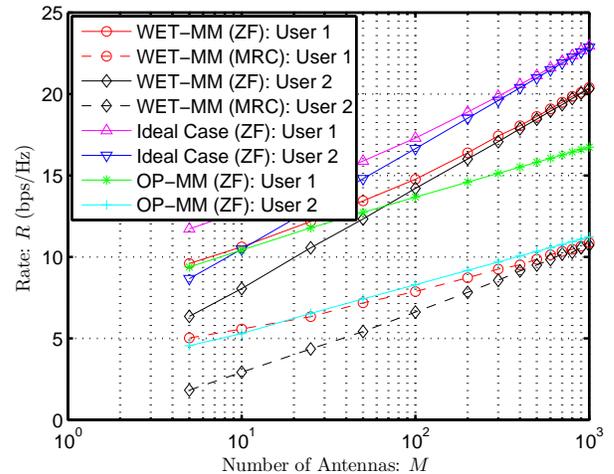}
\caption{Comparison of rate fairness versus $M$.}
\label{fig:Fig5}
\end{figure}

Fig.~\ref{fig:Fig5} compares the user fairness. Using the derived energy allocation weights for downlink WET in~\eqref{eq:coefficientXi}, the two users in the proposed WET-MM system asymptotically achieve the same rate. In contrast, for the OP-MM system, the far user 2 has a much lower rate than the near user 1, even in the massive MIMO regime. Hence, better user fairness is achieved by the proposed WET-MM system.



\section{Conclusion} \label{Conclusion}
This paper studies a WET-enabled massive MIMO system with imperfect CSI. The minimum rate among users is maximized by jointly optimizing the allocation of time and energy resources. The proposed WET-MM system achieves the maximum MM-DoRG achieved by the ideal case with perfect CSI, which is double of the MM-DoRG of the OP-MM system. Moreover, the WET-MM system achieves the best possible fairness among users, since all users asymptotically achieve a common rate. Also, to achieve a desired common rate for all users with a given maximal AP-user distance, the proposed WET-MM system is numerically shown to require fewer antennas at the H-AP, the number of which is roughly a square root of that required by the OP-MM system. The WET-MM system is a promising system to provide high data rate and overcome the energy bottleneck for wireless devices. Nevertheless, some issues related to practical implementation remain to be addressed, such as the effect of antenna correlation, the effect of imperfect channel reciprocity, other fading channel model with line-of-sight, etc.
\appendices
\section{Proof of Lemma~\ref{lemma:AsymOptimalBF}}
\label{AppendixEnergy_generalBF}
Let $\xi_k^{\prime} \in [0,1], \; \theta_i^{\prime} \in [0,1], \; \; \forall k, \; i$, subject to $\sum_{k=1}^K \xi_k^{\prime} + \sum_{i=1}^{M-K} \theta_i=1$. Denote $\bm{\theta}=[\theta_1 \ \theta_2 \ \cdots \ \theta_{M-K}]^T$. Give the channel estimate $\hatbG$, if the beamformer does not have the structure in~\eqref{eq:beamformerMU}, we can write the beamformer as
\begin{align}
  \bw_0(\hatbG)= \sum_{k=1}^K \sqrt{\xi_k^{\prime}} \frac{\hatbg_k}{\| \hatbg_k \|_2} + \sum_{i=1}^{M-K} \sqrt{\theta_i} \bu_i(\hatbG), \label{eq:generalBF}
\end{align}
where $\{\bu_i(\hatbG)\}_{i=1}^{M-K}$ is an orthonormal basis for the orthogonal complement of the space spanned by $\{\hatbg_k\}_{k=1}^K$.
From the property of MMSE estimation, $\hatbg_k$ is independent of $\be_k$. Conditioned on $\hatbg_k$, $\bg_k$ is distributed as $\calC \calN \left(\hatbg_k, \sigma_{e,k}^2 \bI_M\right)$. Hence, we have
\begin{align}
  \bbE_{\bG \left| \hatbG \right.} \big[ \bg_k \bg_k^H \big] = \sigma_{e,k}^2 \bI_M + \hatbg_k \hatbg_k^H. \label{eq:CondMatrix}
\end{align}
With the beamformer in~\eqref{eq:generalBF}, the expected harvested energy by user $k$ is rewritten from~\eqref{EHEnergy0} as
\begin{align}
  &Q_k^{\prime}(L, p_k^{\sf CE}, \alpha, \xi_k^{\prime}, \bm{\theta}) \nonumber \\ 
  &=\alpha p_{\sf dl} \bbE_{\hatbG} \!\left[ \!\bw_0^{H} (\hatbG) \bbE_{\bG | \hatbG} \! \left[\bg_k \bg_k^H \right] \! \bw_0 (\hatbG) \!\right] \nonumber \\
  &\eqa \alpha p_{\sf dl} \bbE_{\hatbG} \Bigg[ \left( \sum_{k=1}^K \sqrt{\xi_k^{\prime}} \frac{\hatbg_k}{\| \hatbg_k \|_2} \right)^H \left( \sigma_{e,k}^2 \bI_M + \hatbg_k \hatbg_k^H \right) \cdot \nonumber \\
  &\quad \left( \sum_{k=1}^K \sqrt{\xi_k^{\prime}} \frac{\hatbg_k}{\| \hatbg_k \|_2} \right) \! \Bigg] \!+\! \alpha p_{\sf dl} \bbE_{\hatbG} \Bigg[\! \left(\sum_{i=1}^{M-K} \! \sqrt{\theta_i} \bu_i(\hatbG) \right)^H \cdot \nonumber \\
  &\quad  \left( \sigma_{e,k}^2 \bI_M + \hatbg_k \hatbg_k^H \right) \left( \sum_{k=1}^K \sqrt{\xi_k^{\prime}} \frac{\hatbg_k}{\| \hatbg_k \|_2} \right) \Bigg] + \alpha p_{\sf dl} \cdot \nonumber \\
  &\quad \bbE_{\hatbG} \Bigg[ \left( \sum_{k=1}^K \sqrt{\xi_k^{\prime}} \frac{\hatbg_k}{\| \hatbg_k \|_2} \right)^H \left( \sigma_{e,k}^2 \bI_M + \hatbg_k \hatbg_k^H \right) \cdot \nonumber \\
  &\quad \left(\sum_{i=1}^{M \!- \!K} \! \sqrt{\theta_i} \bu_i(\hatbG) \! \right) \! \Bigg] \!+\! \alpha p_{\sf dl} \bbE_{\hatbG} \Bigg[ \!\left(\sum_{i=1}^{M \!-\! K} \! \sqrt{\theta_i} \bu_i(\hatbG) \! \right)^H \cdot \nonumber \\
  &\quad \left( \sigma_{e,k}^2 \bI_M + \hatbg_k \hatbg_k^H \right) \left(\sum_{i=1}^{M-K} \sqrt{\theta_i} \bu_i(\hatbG) \right) \Bigg] \nonumber \\
  &\eqb \alpha p_{\sf dl} \xi_k^{\prime} \left[ M \beta_k - (M -1)\sigma_{e,k}^2\right] \nonumber \\
  &\quad + \alpha p_{\sf dl} \beta_k (1-\xi_k^{\prime}) + \sigma_{e,k}^2 \sum \limits_{i=1}^{M-K} \theta_i \label{eq:Energy_generalBF}
\end{align}
where (a) is from~\eqref{eq:generalBF} and~\eqref{eq:CondMatrix}, (b) is from~\eqref{eq:enrgy_optBF} in Appendix~\ref{AppendixEnergy}, as well as the othogonality between $\bu_i(\hatbG)$ and $\hatbg_k$. From the beamformer in~\eqref{eq:generalBF}, we construct a beamformer which has the structure as in~\eqref{eq:beamformerMU} with weights
\begin{align}
  \xi_j = \frac{\xi_j^{\prime}}{\sum \nolimits_{j=1}^K \xi_k^{\prime}}. \label{eq:Relation_Xi}
\end{align}
With the beamformer in~\eqref{eq:beamformerMU} with weights given by~\eqref{eq:Relation_Xi}, the harvested energy is obtained in~\eqref{eq:enrgy_optBF} in Appendix~\ref{AppendixEnergy}.
From~\eqref{eq:Energy_generalBF},~\eqref{eq:Relation_Xi},~\eqref{eq:enrgy_optBF} and~\eqref{eq:Errervar1}, we have
\begin{align}
  Q_k(L, p_k^{\sf CE}, &\alpha, \xi_k) - Q_k^{\prime}(L, p_k^{\sf CE}, \alpha, \xi_k^{\prime}, \bm{\theta}) \nonumber \\
  &\quad=
  \beta_k \sum \limits_{i=1}^{M-K} \frac{\alpha p_{\sf dl} \xi_k^{\prime}(M-1) \beta_k L p_k^{\sf CE} \theta_i }{(\sigma^2+\beta_k L p_k^{\sf CE}) \sum \limits_{j=1}^K \xi_j^{\prime}} \nonumber \\
  &\quad + \beta_k \sum \limits_{i=1}^{M-K} \theta_i \left[ \alpha p_{\sf dl} - \frac{\sigma^2}{\sigma^2+\beta_k L p_k^{\sf CE}}\right] \nonumber \\
  &\quad  \geq 0, \qquad \text{as} \quad M \rightarrow \infty,
\end{align}
where the inequality is from the fact that the $M$-dependent term dominates the minus term. That is, for any beamformer~\eqref{eq:generalBF}, we can always construct the beamformer~\eqref{eq:beamformerMU} to asymptotically harvested more energy. Hence, the asymptotically optimal beamformer has the structure as in~\eqref{eq:beamformerMU}.

\section{Proof of Lemma~\ref{lemma:HarvestedEnergy0}}\label{AppendixEnergy}
Substituting~\eqref{eq:beamformerMU} into~\eqref{EHEnergy0}, the harvested energy is
\begin{align}
  Q_{k} &(L, p_k^{\sf CE}, \alpha, \xi_k) = \alpha \bbE_{\hatbG} \Bigg[ \bbE_{\bG \left| \hatbG \right.} \bigg[ \nonumber \\
   &\quad \frac{p_{\sf dl} \xi_k \left| \bg_k^H \hatbg_k \right|^2}{\| \hatbg_k \|^2} \!+\!
  \frac{p_{\sf dl} \sqrt{\xi_k} \left(\bg_k^H \hatbg_k \right)^H }{\| \hatbg_k \|_2}  \! \sum_{i \neq k} \frac{\sqrt{\xi_i} \bg_k^H \hatbg_i}{\| \hatbg_i \|_2} \nonumber \\
  &\qquad \qquad +\frac{p_{\sf dl} \sqrt{\xi_k} \bg_k^H \hatbg_k} {\| \hatbg_k \|_2}  \sum_{i \neq k} \frac{\sqrt{\xi_i} \left(\bg_k^H \hatbg_i \right)^H}{\| \hatbg_i \|_2} + \nonumber \\
   &\qquad \qquad \sum_{i \neq k} \sum_{j \neq k} \frac{p_{\sf dl} \sqrt{\xi_i \xi_j} \left(\bg_k^H \hatbg_i \right)^H \bg_k^H \hatbg_j}{\| \hatbg_i \|_2 \| \hatbg_j \|_2} \bigg] \Bigg]. \label{EHEnergyA}
\end{align}

In the sequel, we investigate the four terms in~\eqref{EHEnergyA}. Recall $\hatbg_k=\bg_k+\be_k$ and~\eqref{eq:CondMatrix}. Conditioned on the channel estimate $\hatbG$, the harvested energy in the first term is rewritten as
\begin{align}
  \bbE_{\bG \left| \hatbG \right.} \! \left[ \frac{p_{\sf dl} \xi_k \left|  \bg_k^H \hatbg_k \right|^2}{\| \hatbg_k \|^2} \right]  & \!=\! \frac{p_{\sf dl} \xi_k \hatbg_k^H \bbE_{\bg_k \left| \hatbg_k \right.} \left[ \bg_k \bg_k^H \right] \hatbg_k}{\| \hatbg_k \|^2} \nonumber \\
  &=  p_{\sf dl} \xi_k \left( \sigma_{e,k}^2 + \hatbg_k^H \hatbg_k \right) .\label{EHEnergyA1a}
\end{align}

From the fact that $\hatbg_k \sim \calC \calN \left( \mathbf{0}_M, (\beta_k-\sigma_{e,k}^2) \bI_M\right)$, we have that $\bbE_{\hatbg_k} \big[  \hatbg_k^H \hatbg_k \big] = M \left(\beta_k -\sigma_{e,k}^2\right)$.
Hence, the first term in~\eqref{EHEnergyA} is obtained as
\begin{align}
  \alpha \bbE_{\hatbG} \Bigg[ \bbE_{\bG \left| \hatbG \right.} &\left[ \frac{p_{\sf dl} \xi_k \left| \bg_k^H \hatbg_k \right|^2}{\| \hatbg_k \|^2} \right] \Bigg]  \nonumber \\
  &\qquad = \alpha p_{\sf dl} \xi_k \left[ M \beta_k \!-\! (M \!-\!1)\sigma_{e,k}^2\right].\label{EHEnergyA1}
\end{align}
Define $\tilbg_i \triangleq \frac{\hatbg_i}{\| \hatbg_i \|_2}$, and $\bar{\bg}_k \triangleq \| \hatbg_k \|_2 \hatbg_k$. From~\eqref{eq:CondMatrix}, the second term in~\eqref{EHEnergyA} is rewritten as
\begin{align}
  \alpha &\bbE_{\hatbG} \left[ \bbE_{\bG \left| \hatbG \right.} \left[ \frac{ p_{\sf dl} \sqrt{\xi_k} \left(\bg_k^H \hatbg_k \right)^H }{\| \hatbg_k \|_2}  \sum_{i \neq k} \frac{\sqrt{\xi_i} \bg_k^H \hatbg_i}{\| \hatbg_i \|_2}\right] \right]
  \nonumber \\
   &\quad = \alpha {p_{\sf dl} \sqrt{\xi_k} \sigma_{e,k}^2 }\sum_{i \neq k} {\sqrt{\xi_i} \bbE_{\tilbg_k, \tilbg_i} \big[ \tilbg_k^H \tilbg_i \big] } \nonumber \\
   &\qquad + \alpha {p_{\sf dl} \sqrt{\xi_k}}  \sum_{i \neq k} {\sqrt{\xi_i} \bbE_{\bar{\bg}_k, \tilbg_i} \big[ \bar{\bg}_k^H \tilbg_i \big]} \; \eqa 0,  \label{EHEnergyA2}
\end{align}
where ($a$) is from the fact that $\tilbg_k$ and $\bar{\bg}_k$ are independent zero-mean random vectors, for any $i \neq k$.

The third term in~\eqref{EHEnergyA}, which is the conjugate of the second term in~\eqref{EHEnergyA}, is similarly obtained as
\begin{align}
  \! \alpha \bbE_{\hatbG} \! \left[ \bbE_{\bG \left| \hatbG \right.} \! \left[ \frac{ p_{\sf dl} \sqrt{\xi_k} \bg_k^H \hatbg_k }{\| \hatbg_k \|_2}  \sum_{i \neq k} \frac{\sqrt{\xi_i} \left( \bg_k^H \hatbg_i \right)^H}{\| \hatbg_i \|_2}\right] \right] & \! = \! 0. \label{EHEnergyA3}
\end{align}

The fourth term in~\eqref{EHEnergyA} is rewritten as

\begin{align}
  \alpha \bbE_{\hatbG} &\left[ \bbE_{\bG \left| \hatbG \right.} \left[ \sum_{i \neq k} \sum_{j \neq k} \frac{p_{\sf dl} \sqrt{\xi_i \xi_j} \left(\bg_k^H \hatbg_i \right)^H \bg_k^H \hatbg_j}{\| \hatbg_i \|_2 \| \hatbg_j \|_2} \right] \right] \nonumber \\
  &\quad \eqa \alpha p_{\sf dl} \bbE_{\hatbG} \big[ \sum_{i \neq k} {\xi_i \tilg_i^H (\sigma_{e,k}^2 \bI_M + \hatbg_k \hatbg_k^H) \tilbg_i} \nonumber \\
  &\quad \quad + \sum_{i \neq k} \sum_{j \neq i, k} {\sqrt{\xi_i \xi_j} \tilbg_i^H (\sigma_{e,k}^2 \bI_M + \hatbg_k \hatbg_k^H) \tilbg_j}\big] \nonumber \\
  &\quad \eqb \alpha p_{\sf dl} \beta_k \sum_{i \neq k} \xi_i , \label{EHEnergyA4}
\end{align}
where ($a$) is from~\eqref{eq:CondMatrix}, and ($b$) is from the fact that $\bbE_{\hatbG} \left(\hatbg_k \hatbg_k^H\right)=(\beta_k - \sigma_{e,k}^2) \bI_M$, and $\tilbg_i$ and $\tilbg_j$ are independent zero-mean random vectors, for any $i \neq j$.

Substituting~\eqref{EHEnergyA1},~\eqref{EHEnergyA2},~\eqref{EHEnergyA3} and~\eqref{EHEnergyA4} into~\eqref{EHEnergyA}, we obtain the harvested energy as
\begin{align}
  Q_k(L, p_k^{\sf CE}, \alpha, \xi_k) &= \alpha p_{\sf dl} \xi_k \left[ M \beta_k \!-\! (M \!-\!1)\sigma_{e,k}^2\right] \nonumber \\
  &\quad + \alpha p_{\sf dl} \beta_k (1-\xi_k). \label{eq:enrgy_optBF}
\end{align}
Substituting~\eqref{eq:Errervar1} into~\eqref{eq:enrgy_optBF}, the harvested energy is in~\eqref{HarvestedEnergy0}.

\section{Proof of Lemma~\ref{LemmaMRC}}\label{AppendixMRC}
With MRC, $\ba_k=\hatbg_k$. Define $\tilg_i \triangleq \frac{ \hatbg_k^H \hatbg_i}{\| \hatbg_k \|}$.
The achievable rate of user $k$ is written as
\begin{align}
  &\tilR_k^{\text{MRC}} =(1-\tau-\alpha) \nonumber \\
  &\log \left( 1+ \left(\bbE \left\{ \frac{\sum \limits_{i=1, i \neq k}^K p_i |\tilg_i|^2 + \sum \limits_{i=1}^K p_i \sigma_{e,i}^2 + \sigma^2 }{p_k  |\hatbg_k^H \hatbg_k |} \right\} \right)^{-1} \right). \label{eq:RateLBMRC1}
\end{align}
It can be easily shown that $\tilg_i \sim \calC \calN (0, \beta_i - \sigma_{e,i}^2)$. It is further noted that the conditional probability density function (pdf) $f(\tilg_i | \hatbg_k) = f(\tilbg_i)$, where $f(\tilbg_i)$ is the marginal pdf of $\tilg_i$. Then we have
\begin{align}
  &\bbE \left\{ \frac{\sum \limits_{i=1, i \neq k}^K p_i |\tilg_i|^2 + \sum \limits_{i=1}^K p_i \sigma_{e,i}^2 + \sigma^2 }{p_k  |\hatbg_k^H \hatbg_k |} \right\} = \nonumber \\
  &\left( \sum \limits_{i=1, i \neq k}^K p_i \beta_i + p_k \sigma_{e,k}^2 + \sigma^2  \right) \bbE \left\{ \frac{1}{p_k  |\hatbg_k^H \hatbg_k |} \right\}. \label{eq:RateLBMRC2}
\end{align}

It can be shown that the random variable $Z_k \triangleq \frac{2}{\beta_k-\sigma_{e,k}^2} \hatbg_k^H \hatbg_k$ follows central chi-square distribution with $2M$ degrees of freedom. Then, we have $ \bbE \left( \frac{1}{Z_k} \right) = \frac{1}{2(M-1)}$. Hence, it holds that
\begin{align}
  \bbE \left\{ \frac{1}{p_k  |\hatbg_k^H \hatbg_k |} \right\} &= \frac{2}{p_k (\beta_k-\sigma_{e,k}^2)} \bbE \left\{ \frac{1}{Z_k} \right\} \nonumber \\
  &= \frac{1}{p_k (M-1) (\beta_k-\sigma_{e,k}^2)}.\label{eq:RateLBMRC3}
\end{align}
Substituting~\eqref{eq:RateLBMRC2} and~\eqref{eq:RateLBMRC3} into~\eqref{eq:RateLBMRC1}, we obtain that
\begin{align}
      &\tilR_k^{\text{MRC}} = (1-\tau-\alpha) \nonumber \\
       &\log \left( \! 1 \!+\! \frac{p_k (\tau, \alpha, \rho) (M-1) \big( \beta_k-\sigma_{e,k}^2 (\alpha, \rho) \big)}{\sum \limits_{ i \neq k} p_i (\tau, \alpha, \rho) \beta_i + p_k (\tau, \alpha, \rho) \sigma_{e,k}^2 (\alpha, \rho) + \sigma^2} \right). \label{eq:RateLBMRC0}
\end{align}
The result in~\eqref{eq:RateLBMRC} is obtained, by substituting the error variance in~\eqref{ErrorVariance} into~\eqref{eq:RateLBMRC0}.

\section{Proof for Lemma~\ref{Lemma:CoeffDesign}}\label{AppendixCoeff}
We rewrite the asymptotic achievable rate in~\eqref{eq:RateLBZFAsymp1} for ZF detection as follows,
\begin{align}
    \tilR_k^{\text{ZF}} &\convergeM  (1-\tau-\alpha) \log \big( 1+ C_k(\tau, \alpha, \rho) \xi_k \big), \label{eq:RateLBMRCAsymp2}
  \end{align}
where $C_k(\tau, \alpha, \rho)$ is the multiplicative coefficient of $\xi_k$ in the logarithm of~\eqref{eq:RateLBZFAsymp1}.

%
We consider the case in which the first $(K-1)$ users have the same rate $\tilR^{\text{ZF}}$ and the $K$-th user achieves a higher rate, i.e., $\tilR_1^{\text{ZF}}=\tilR_2^{\text{ZF}}=\cdots=\tilR_{K-1}^{\text{ZF}} =\tilR^{\text{ZF}} < \tilR_K^{\text{ZF}}$. Clearly, one can always increase the minimum rate $\tilR^{\text{ZF}}$ by increasing $\xi_1, \xi_2,\cdots, \xi_{K-1}$ and decreasing $\xi_K$, subject to the constraint $\sum_{i=1}^K \xi_i=1$. Then the minimum rate among users is maximized when $\tilR^{\text{ZF}} = \tilR_K^{\text{ZF}}$.

The same argument can be extended to other cases where less than $(K-1)$ users have the same UL rate. That is, the minimum rate is maximized when all users achieves the same rate; otherwise, one can always increase the minimum rate by adjusting $\xi_k$'s. Let $\tilR^{\text{ZF}}=(1-\tau-\alpha) \log ( 1+ \gamma )$, where the common SINR $\gamma = C_k(\tau, \alpha, \rho) \xi_k$. From the constraint $\sum_{i=1}^K \xi_i=1$, we obtain that
\begin{align}
\xi_k^{\star}&= \frac{\gamma}{C_k(\tau, \alpha, \rho)} \nonumber \\
&= \frac{1}{C_k(\tau, \alpha, \rho) \sum_{i=1}^K \frac{1}{C_i (\tau, \alpha, \rho)}}= \frac{{1}}{\beta_k^2 \sum \nolimits_{i=1}^K \frac{1}{\beta_i^2}}. \nonumber
\end{align}
Hence, the asymptotically optimal $\xi_k^{\star}$ only depends on the long-term path loss $\beta_i$'s of all users.

For MRC detection, by using the Karush-Kuhn-Tucker conditions~\cite{ConvecOptBoyd04}, it can be shown that the minimum rate is maximized when all users achieves the same rate. The energy allocation weight $\xi_k^{\star}$ is thus given by~\eqref{eq:coefficientXi}.

\bibliography{IEEEabrv,reference1312}
\bibliographystyle{ieeetr}

\end{document}